\documentclass[
 reprint, 			
 superscriptaddress,
 nofootinbib,
 amsmath,amssymb,
 aps,
 pra,
 floatfix,
]{revtex4-1}

\usepackage{graphicx}  
\usepackage{dcolumn}   
\usepackage{bm}        
\usepackage{gensymb}
\usepackage{float}	   
\usepackage{scrextend}
\usepackage{soul}
\usepackage{array}
\usepackage{booktabs,siunitx}
\usepackage[caption=false]{subfig}  
\usepackage{cleveref}
\usepackage{url}
\begin{document}


\title{Search for Perturbations of Nuclear Decay Rates \\Induced by Reactor Electron Antineutrinos}


\author{V.E. Barnes}
\affiliation{Department of Physics $\&$ Astronomy,\\Purdue University, West Lafayette, IN 47907 U.S.A.}
\affiliation{SNARE Inc., West Lafayette, IN 47906 U.S.A.}

\author{D.J. Bernstein}
\affiliation{Department of Physics $\&$ Astronomy,\\Purdue University, West Lafayette, IN 47907 U.S.A.}

\author{C.D. Bryan}
\affiliation{High Flux Isotope Reactor (HFIR),\\Oak Ridge National Laboratory, P.O. Box 2008, Oak Ridge, TN 37831 U.S.A.}

\author{N. Cinko}
\affiliation{Department of Physics $\&$ Astronomy,\\Purdue University, West Lafayette, IN 47907 U.S.A.}

\author{\\G.G. Deichert}
\affiliation{High Flux Isotope Reactor (HFIR),\\Oak Ridge National Laboratory, P.O. Box 2008, Oak Ridge, TN 37831 U.S.A.}

\author{J.T. Gruenwald}
\affiliation{Department of Physics $\&$ Astronomy,\\Purdue University, West Lafayette, IN 47907 U.S.A.}
\affiliation{SNARE Inc., West Lafayette, IN 47906 U.S.A.}

\author{J.M. Heim}
\affiliation{Department of Physics $\&$ Astronomy,\\Purdue University, West Lafayette, IN 47907 U.S.A.}

\author{H.B. Kaplan}
\affiliation{Department of Physics $\&$ Astronomy,\\Purdue University, West Lafayette, IN 47907 U.S.A.}

\author{\hbox{R. LaZur}}
\affiliation{Department of Physics $\&$ Astronomy,\\Purdue University, West Lafayette, IN 47907 U.S.A.}

\author{D. Neff}
\affiliation{Department of Physics $\&$ Astronomy,\\Purdue University, West Lafayette, IN 47907 U.S.A.}

\author{J.M. Nistor}
\affiliation{Department of Physics $\&$ Astronomy,\\Purdue University, West Lafayette, IN 47907 U.S.A.}

\author{N. Sahelijo}
\affiliation{Department of Physics $\&$ Astronomy,\\Purdue University, West Lafayette, IN 47907 U.S.A.}

\author{E. Fischbach}
\thanks{Corresponding Author: ephraim@purdue.edu}
\affiliation{Department of Physics $\&$ Astronomy,\\Purdue University, West Lafayette, IN 47907 U.S.A.}
\affiliation{SNARE Inc., West Lafayette, IN 47906 U.S.A.}


\date{\today}


\begin{abstract}
We report the results of an experiment conducted near the High Flux Isotope Reactor of Oak Ridge National Laboratory, designed to address the question of whether a flux of reactor-generated electron antineutrinos ($\overline{\nu}_e$) can alter the rates of weak nuclear interaction-induced decays for $^{54}$Mn, $^{22}$Na, and $^{60}$Co.  This experiment, while quite sensitive, cannot exclude perturbations less than one or two parts in 10$^4$ in $\beta^{\pm}$ decay (or electron capture) processes, in the presence of an antineutrino flux of \hbox{3$\times$10$^{12}$ cm$^{-2}$s$^{-1}$}.  The present experimental methods are applicable to a wide range of isotopes.  Improved sensitivity in future experiments may be possible if we can understand and reduce the dominant systematic uncertainties. 
\begin{description}
\item[Keywords]
Reactor Physics, Neutrino Physics
\end{description}
\end{abstract}


\maketitle                  

\section{Introduction} \label{Introduction}

Few issues frame the history of natural radioactivity as fundamentally as the question of whether radioactive decays are unaffected by their local environment. Although there is compelling evidence to support this, recent studies have reported evidence of a solar influence on certain radioactive decay processes. This includes annual oscillations \cite{1,2,3,4,5,6,7,8,9,10,11,12,13,14,15,16,17,18,19,20,21,22,23} and indications of frequencies associated with solar rotation \cite{14,16}. Additionally, a suggestion for a possible solar influence on nuclear decays comes from observations of short-term changes associated with solar storms \cite{24,25}. Although some questions have been raised concerning the data supporting a solar influence \cite{26,27,28,29}, they have been addressed in the literature \cite{30,21,13}. As noted in Table II of Ref. \cite{21}, the indications of a possible solar influence on radioactive decays come from experiments using a variety of detectors monitoring a number of different isotopes. One hypothesis which could account for these observations is that they are due to the influence of solar neutrinos through some as yet unknown mechanism. This motivates the study of decay rates in the presence of more readily available electron antineutrinos \cite{lindstrom1,lindstrom2}, including those produced by nuclear reactors.\\

It is now well established that there exist three types of neutrinos denoted by $\nu_e$, $\nu_\mu$, $\nu_\tau$ and their corresponding antiparticles $\overline{\nu}_e$, $\overline{\nu}_\mu$, $\overline{\nu}_\tau$. Additional types of neutrinos or neutrino-like particles may exist such as sterile neutrinos \cite{sterileneutrinos} and neutrellos \cite{neutrellos}. In what follows, we describe a reactor experiment aimed at studying whether nuclear decay rates are influenced by any light, neutral, and weakly interacting particle emitted by a reactor, including (but not limited to) the presumably dominant $\overline{\nu}_e$. We note that the limits presented below on perturbations induced by ̅$\overline{\nu}_e$ do not necessarily apply to the other neutrino flavors $(\overline{\nu}_\mu,\overline{\nu}_\tau,...)$ which may have significantly different properties (e.g. magnetic moments).\\

Our experiment was carried out in two phases at the 85 MW (thermal) High Flux Isotope Reactor (\textsc{hfir}), located at Oak Ridge National Laboratory in Oak Ridge, Tennessee. This choice was motivated in part by the opportunity to position our experiment sufficiently close to the reactor core in order to achieve a $\overline{\nu}_e$ flux comparable to, or larger than, the solar neutrino flux ($\nu_\odot$). Additionally, the routine reactor refueling outages introduce a convenient step function in neutrino flux as the driving signal in our spectrometers. Planning for the experiment began in October 2013, and our first run began in March 2014. Since initiating our reactor experiment at \textsc{hfir}, we learned of an alternate reactor experiment by de Meijer and Steyn \cite{deMeijer2} which improves on an earlier experiment by de Meijer, Blaauw, and Smit \cite{deMeijer1}. Although the earlier paper found a null result, the more recent paper now reports a positive effect, which the authors attribute to either an interaction between $\overline{\nu}_e$ and their $^{22}$Na sample, or else to a hidden instrumental cause.\\ 

Phase I of this experiment was an initial exploratory period which focused on understanding various systematic effects such as backgrounds and the sensitivity of our detectors to environmental conditions.  Our experimental results are derived from Phase II of this experiment, lasting from August of 2014 to March of 2015.  Due to its extended duration of 217 days, with five reactor-\textsc{on} periods and more accurate temperature control, Phase II achieved at least an order of magnitude better sensitivity than Phase I.

\section{Phase I Experiment}
Data acquisition in Phase I commenced on 15 March 2014 at 04:24 with the reactor \textsc{on} and continued through the reactor shutdown (23 March 2014 at 05:23) until 30 March 2014.  This provided for an \textsc{on} / \textsc{off} comparison of the decay rates of the three radioactive isotopes that were studied: $^{54}$Mn which decays by electron capture (EC), $^{60}$Co which decays by beta-emission ($\beta^-$), and $^{152}$Eu which has significant branching to both EC (72.10$\%$) and $\beta^-$ decay ($27.90\%$), see Table \ref{tab:decaychains} of Appendix \ref{app:sources}. The three radioactive samples studied in this experiment were chosen, in part, from indications in earlier experiments \cite{1,2,3,4,5,6,7,8,9,10,11,12,13,14,15,16,17,18,19,20,21,22,23} that each had exhibited a time variation in its decay parameter, possibly due to solar neutrinos.\\ 

Further motivation for the choice $^{54}$Mn comes from an apparent correlation between a change in the $^{54}$Mn decay rate and the solar storm of 13 December 2006 \cite{24}, and subsequent storms \cite{25}. This isotope is also of interest since its dominant decay mode is electron capture, and hence allows a comparison to the isotope $^{60}$Co, which a pure $\beta^-$ decay. Cobalt-60, in turn, was selected on the basis of the observation by Parkhomov \textit{et al}. of an annual variation in its decay \cite{parkhomovannual}. Finally, the choice of $^{152}$Eu was dictated by several considerations including the fact that this isotope decays via both EC and $\beta^-$ modes as noted above. Moreover, the data presented by Seigert \textit{et al}. \cite{3} indicate the presence of a periodic signal arising from the 1408 keV $\gamma$ photon emitted in the EC process. This particular observation is especially interesting since the periodic signal, obtained using a Ge(Li) detector, was observed to be ∼180 days out of phase with the data taken by the same group on $^{226}$Ra and its daughters using a $4\pi$  $\gamma$-ionization chamber.\\

Additionally, isotopes were preferred which decayed to the excited state of the daughter nucleus. This would ensure that the $\beta-$decay of the parent could eventually be recorded by the photons emitted in the nuclear de-excitation of the daughter to its ground state. Since such a decay produces a sharp peak at a well-defined energy, this allowed us to focus on a relatively narrow region-of-interest (\textsc{roi}) in the $\gamma$-spectrum whose location in a Multi-Channel Analyzer (MCA) could be identified and controlled. Among isotopes whose decays suggest interesting effects, $^{3}$H, $^{32}$Si, and $^{36}$Cl are decays with no photons, and thus not suited to be detected by our NaI crystal scintillators. These considerations led to our selection of $^{54}$Mn, $^{60}$Co, and $^{152}$Eu in Phase I, joined by $^{22}$Na in Phase II. \\

Particular attention was devoted to $^{54}$Mn (T$_{1/2}$ = 312d) with which we had the most extensive experience from our previous experiments at Purdue. $^{54}$Mn decays via electron capture and is detected via the chain :\\

\begin{flushleft} \label{eq:A.1}
$^{54}_{25}$Mn + $e^-$(EC) $\rightarrow$ $^{54}$Cr$^*$ + $\nu_e$ (100$\%$) \\

\smallskip
\hspace{3.2cm}$\hookrightarrow^{54}_{24}$Cr(g.s.) + $\gamma$(834.8 keV) \\

\end{flushleft} 

We note in passing that our choices led to a set of isotopes with a broad range of half-lives: $^{54}$Mn (\hbox{312 d}), $^{22}$Na (\hbox{2.60 a}), $^{60}$Co (\hbox{5.27 a}), $^{152}$Eu (\hbox{13.54 a}).  All published half-lives are taken from Ref. \cite{half-lives}.  The sources have activities of several $\mu$Ci (1 Ci = 3.7$\times$10$^{10}$ decays per second) and give counting rates of order 20 to 35 kHz, which entail detector dead-times in the range 10$\%$ to 13$\%$. The sources were encapsulated in standard 1-inch diameter plastic discs.\\

In Phase I of this experiment the apparatus consisted of four detectors, including one with no source mounted as a control and to measure background counting rates. The detectors were St. Gobain/\textsc{bicron} 2 inch NaI(Tl) crystal plus photomultiplier, sealed assemblies, in 2.25 inch diameter cylindrical aluminum cans. Each detector was coupled to an \textsc{ortec} digi\textsc{base}, an electronically gain-locked PMT base, including high voltage supply, preamplifier, and multichannel analyzer. Dead-time in the system was corrected by implementation of a variant of the Gedcke-Hale live-time clock, which extended the running time of a measurement cycle until the requested live-time was achieved. The detectors were each attached, by one USB to a PC running \textsc{ortec maestro} software which provided an interface to communicate to the detectors. \\

The source discs were each mounted on a triangular metal plate affixed to three thin “standoffs” matching the periphery of the front end of the detector can. These standoffs were then affixed to the edge of the front face of the can. The objective is to avoid motion of the source by possible barometric flexing of the front disc of the sealed can. (This had been seen to happen if the source is mounted directly to the center of the front disc, with small, concomitant but significant fluctuations in counting rate.) The objective of the experiment was to be sensitive to fluctuations in counting rates smaller than 10$^{-3}$.  This requires very stable geometry of the source position relative to the detector: a simple geometrical calculation shows that a point-like source located 6 mm from a 50 mm diameter, 50 mm long crystal (typical of our assemblies), and displaced 1 mm towards or away from the crystal, sees a change in the subtended solid angle of $\pm4.8 \%$ at the front face of the crystal, and $\pm3.14 \%$ at the back face of the crystal.  This represents an average of $\pm4\%$ per 1 mm motion, or as a useful mnemonic, 1 part per thousand per 0.001 inches (25 $\mu$ m) of motion.\\

\begin{figure}
	\subfloat[]{\includegraphics[width=0.48\textwidth]{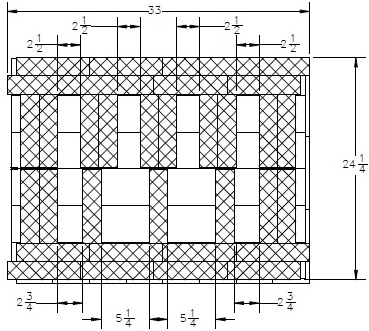}}
    
    \subfloat[]{\includegraphics[width=0.48\textwidth]{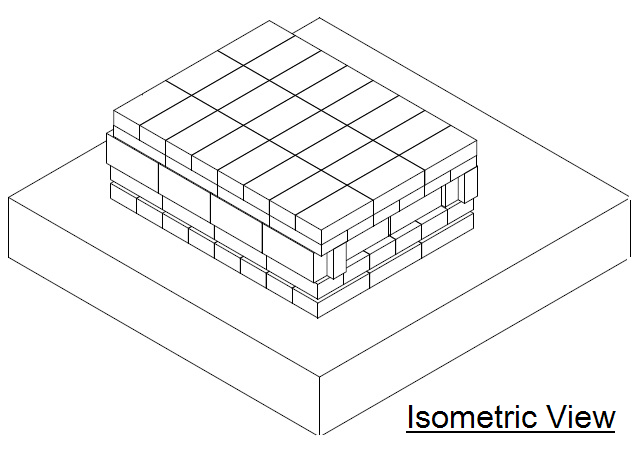}}
     \caption{(a) Schematic of the layout of each detector bay within the lead cave (b) Diagram of the lead cave in Phase I}   
    \label{fig:detectorbay}
    
\end{figure}

Background suppression was achieved by surrounding each detector by at least 4 inches (10 cm) of lead. Moreover, the detectors were separated from one another by 4 inches of lead to mitigate cross-talk in a structure, referred to as the cave, with four bays (see Fig. \ref{fig:detectorbay} for details). Altogether, approximately 1.8 metric tons of lead were required, using 150 standard 2$\times$4$\times$8 inch$^3$ lead bricks. \\

With the reactor \textsc{on}, backgrounds outside the cave were typically a few thousand counts per second (cps). The lead shielding was effective in reducing the background by nearly 3 orders of magnitude, and we conservatively mounted the background detector in the highest-count bay (bay $\#$1). The gain of the background counter was set lower than any of the other detectors, so that it was sensitive to any gamma ($\gamma$) energy that would arise from any of our isotopes. The above counting rate refers to the entire spectral range of the background counter, whereas only portions of the background will fall in the regions of interest (\textsc{roi}s) used for counting.\\

The overall counting rates from backgrounds inside the cave were small, of order one part in one thousand of the counting rates from the sources. Hence in Phase I, background estimates were not subtracted from the source counts. Moreover, fluctuations in the interior background rates were less than 20$\%$ of the average background level, and were believed to be associated with the intermittent operation of a neutron beam line passing one floor below in the reactor building. Changes in background rates were considered to be negligible for the exploratory purposes of Phase I.\\ 

The lead cave, and the detector array, were situated \hbox{5.83 m} from the \textsc{hfir} reactor core at site EF-4, directly along one of the reactor-core center-lines: 4 m higher than the core and 4.27 m away laterally. The core is a cylinder 20 cm in diameter and 40 cm in height, which we treat approximately as point-like for flux calculations. Standard calculations of the total $\overline{\nu}_e$ flux from highly enriched uranium reactors give a $\overline{\nu}_e$ flux at 5.8 m from \textsc{hfir}, at 85 MW thermal output, of approximately 3.8$\times$10$^{12}$  cm$^{-2}$ s$^{-1}$, which is 58 times larger than the solar $\nu$ flux at Earth, 6.5$\times$10$^{10}$  cm$^{-2}$ s$^{-1}$, and 830 times larger than the 7$\%$ annual variation in the solar $\nu$ flux due to the Earth's orbital eccentricity.\\

\begin{figure}
	\includegraphics[width=0.48\textwidth]{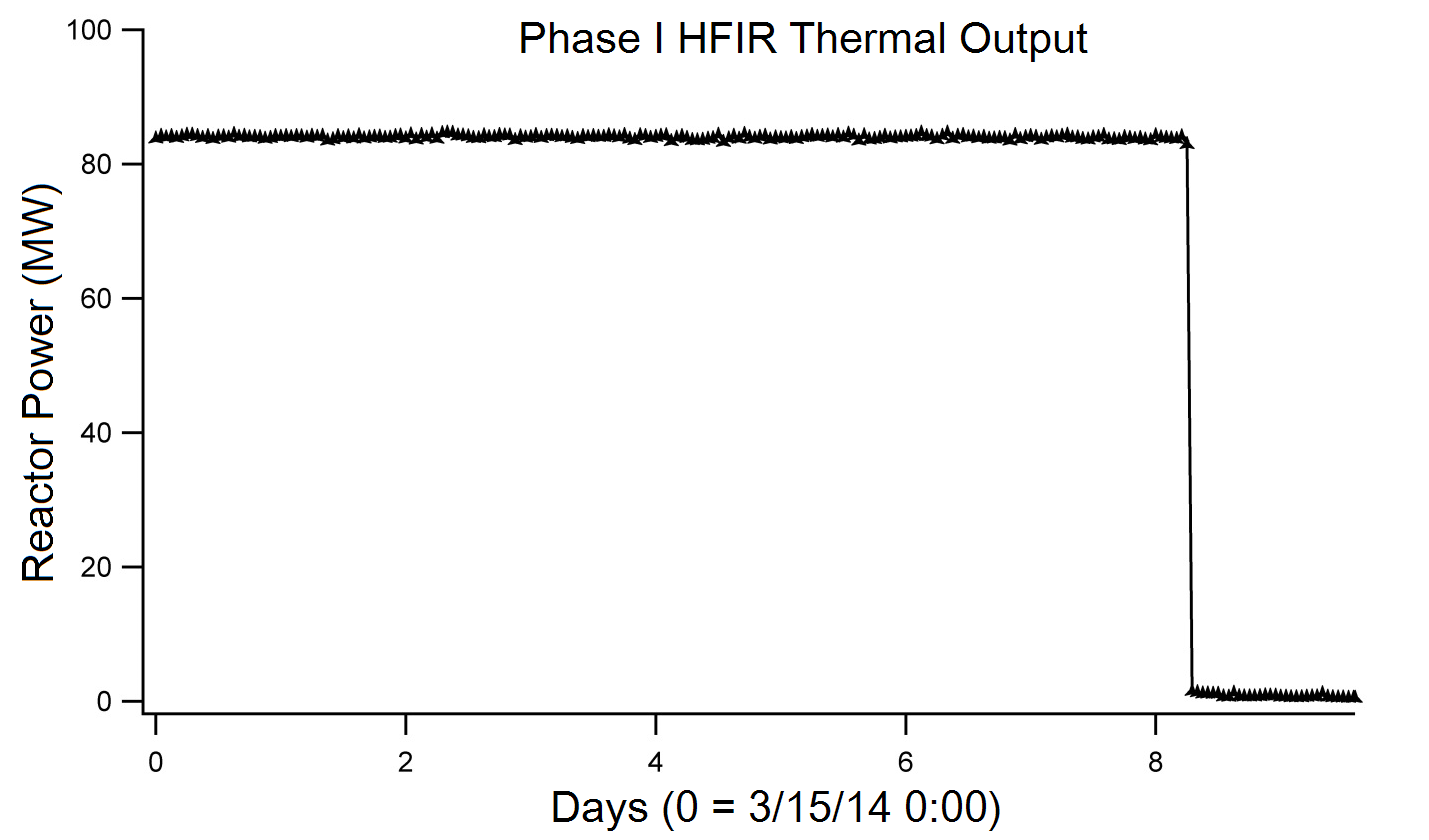}
     \caption{\textsc{hfir} thermal power (MW) output for the first 10 days of the Phase I experiment. The \textsc{hfir} reactor was shut down for routine maintenance on 30 March 2014, at 06:00, which corresponds to day 8.25 on the above graph.} \label{fig:HFIRthermal}
   
\end{figure}

As noted above, primary data acquisition started on 15 March 2014 at 04:24 EDST with the reactor already \textsc{on}, and consisted of a continuous stream of 15 minute (live-time) cycles. Each cycle produced a $\gamma$ energy spectrum, and a table of counts from each of a set of selected \textsc{roi}s (i.e. groups of bins in the energy spectrum). The reactor ran at approximately 85 MW thermal power (see Fig. \ref{fig:HFIRthermal}) until it was shut down at 05:23 on 23 March 2014. Thermal power dropped to 0.2 MW in the succeeding three hours and then effectively to zero, presumably reflecting the initial part of the afterglow arising from short-lived fission products. Data acquisition then continued in the \textsc{off} period until 30 March 2014.\\

Residual $\overline{\nu}_e$ ̅flux from the spent core is expected to be below 1$\%$. Although multiple spent cores are stored at the other end of the reactor pool in which the reactor operates, they are at distances from our detectors several times the distance of the operating reactor. The most recently extracted core emits copious Cherenkov radiation as seen from the observation gallery, however the next most recent core shines only dimly.  The ages of the spent cores increase in increments of roughly two months. Suppression of $\overline{\nu}_e$ flux due to decay time and distance were estimated to outweigh the increase in $\overline{\nu}_e$ flux due to the number of spent cores stored in the pool, as seen at our detectors. \\

\begin{figure}
	\includegraphics[width=0.48\textwidth]{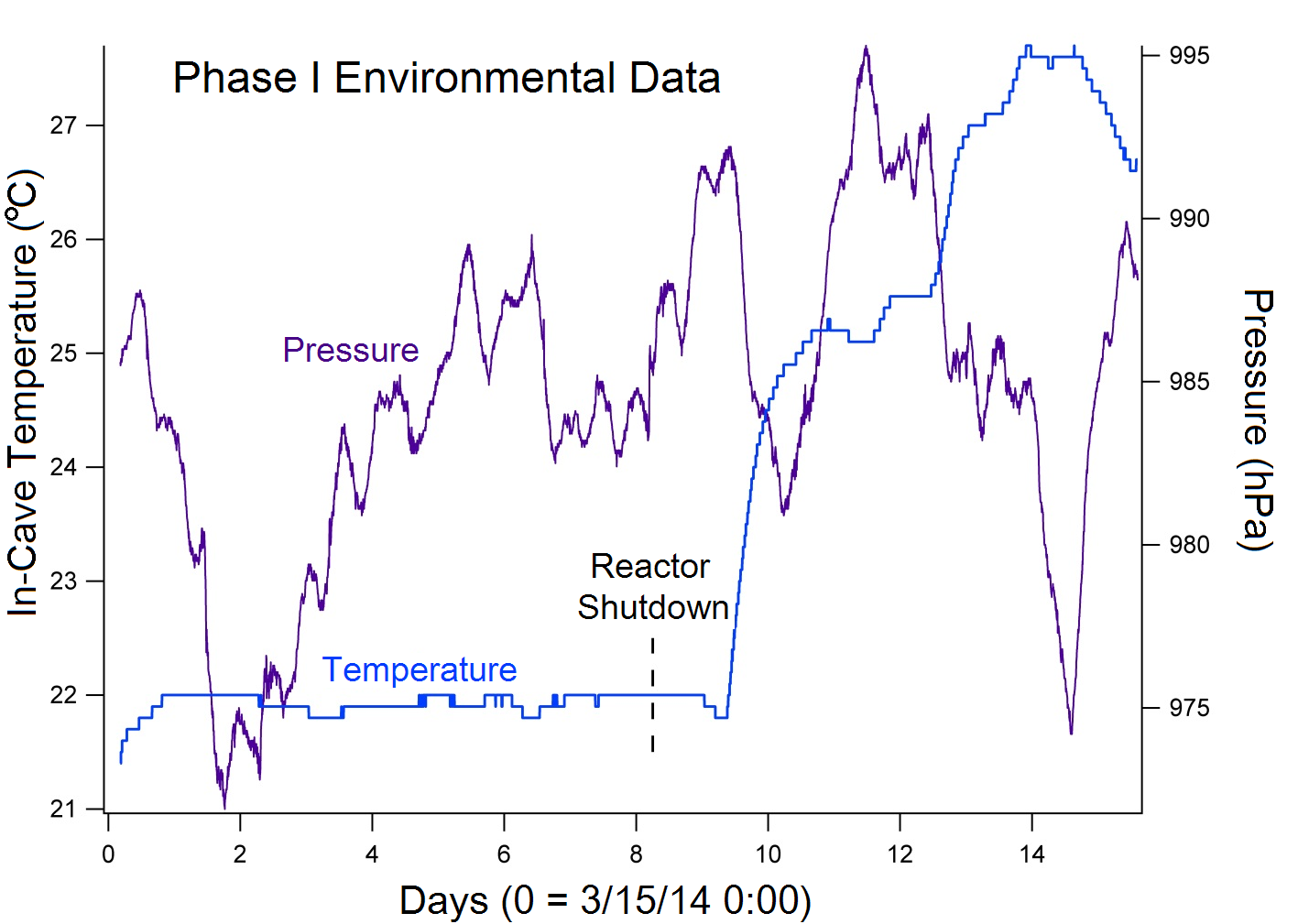}
     \caption{Plot depicting temperature and pressure data as recorded in bay $\#$4 for the first 15 days of the experimental run. The ambient temperature in \textsc{hfir} demonstrated remarkable stability during the reactor-\textsc{on} phase. After reactor shutdown (day 8.25), significant fluctuations in the temperature were recorded.} \label{fig:bay4temppress}  
\end{figure}

Temperature control of the detector-source assembly is very important. Differential thermal expansion of the NaI(Tl) crystal mounted onto a glass PMT envelope, versus the surrounding aluminum can, will partly but not necessarily completely cancel shifts of the front end of the can relative to the crystal. Sodium iodide has a larger expansion coefficient, and glass has a lower coefficient, compared to aluminum. Also, expansion of the NaI(Tl) crystal will increase the solid angle subtended. The temperature at the experimental location was climate controlled within approximately 2 $\degree$C and, as monitored at one of the detectors inside the lead cave, remained within a 0.2 $\degree$C range for most of the period while the reactor was on. This climate control in the hall was unexpectedly (due to HVAC failure) discontinued when the reactor was shut down, following which the temperature measured by the sensor located at bay $\#$4 in the lead cave rose by 6 $\degree$C from day 9 to day 15, in two stages, and then started to drop (see Fig. \ref{fig:bay4temppress}). This motivated incorporation of an independent temperature control for the Phase II experiment. \\

\section{Phase I Results}

The \textsc{roi}s used are shown in the spectral plots of Fig. \ref{fig:energyspectra} in Appendix \ref{app:sources}. For $^{60}$Co the measured half-lives are obtained from both of the prominent $\gamma$-peaks. In the case of $^{152}$Eu, we quote the results obtained from the 3 dominant \textsc{roi}’s, noting that \textsc{roi} $\#$5 arises from the $\beta$-decay mode, whereas \textsc{roi} $\#$1 and \textsc{roi} $\#$2 arise from the EC mode. Table \ref{tab:phase1fits} presents our half-life results for each isotope for the “reactor-\textsc{on}” period, along with the published values obtained from \cite{half-lives}.  We see from this table that during the reactor-\textsc{on} mode, the half-lives of $^{152}$Eu and $^{54}$Mn  were found to be systematically lower than their respective published half-lives, corresponding to higher decay rates. We emphasize that the errors quoted in Table \ref{tab:phase1fits} are purely statistical, and will necessarily be modified by the inclusion of additional uncertainties arising from systematic effects to be discussed below. That such effects may be present is suggested by the values of $\chi^2$/dof (degrees of freedom) in the final column of Table \ref{tab:phase1fits}. Although these are generally close to the “ideal” value of unity, deviations from unity are also evident, which may be connected to some of the effects we discuss below. \\

In contrast to the data obtained when the reactor was in the \textsc{on} mode (i.e. pre-shutdown), which were fairly stable, the data with the reactor in the \textsc{off} mode (i.e. post-shutdown) present a number of challenges and puzzles. Since the reactor-\textsc{on} half-lives are systematically smaller than the published values, we expected these to return fairly quickly to the published values shown in Appendix \ref{app:sources}, assuming a real effect. However, we see from Table \ref{tab:phase1fits} that this is not shown by the data.\\

\begin{figure}
	\includegraphics[width=0.48\textwidth]{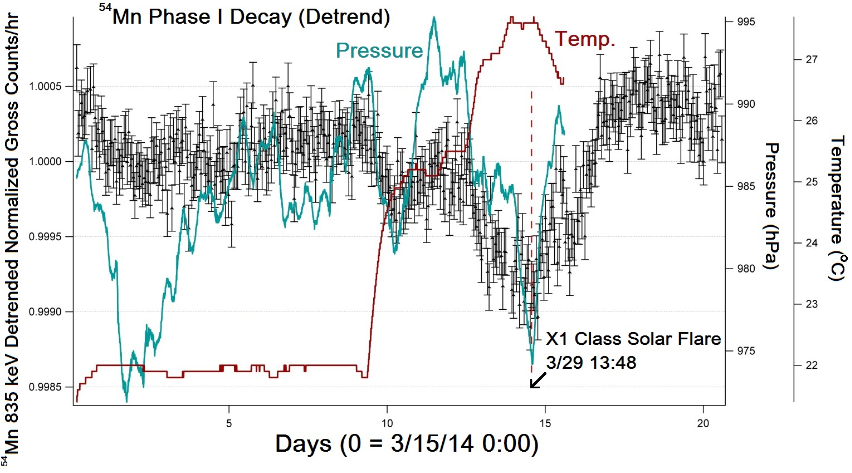}
     \caption{Plot of the T$_{1/2}$-detrended counts recorded in 1 hr bins vs. time for dominant $^{54}$Mn peak during the first 21 days. The reactor shutoff occurred at day 8.25, and soon thereafter a significant rise in temperature was recorded (red curve). The significant “dip” in counts appears to coincide with this temperature change, as well as, with a large X1 Class Solar Flare.}	\label{fig:detrendmn}
    \centering
\end{figure}

On 29 March 2014 our data exhibited a precipitous drop in the count rate of our $^{54}$Mn sample, which coincided with a spike in the recorded temperature as shown in Fig. \ref{fig:detrendmn}. These events also coincided in time with an X1 solar flare which occurred at 13:48 local time (See Fig. \ref{fig:detrendmn}). During the same period, the minimum of the dip in $^{60}$Co count rates also coincided with the solar flare, and the start of the dip preceded the flare by about one day. The two lower-energy $^{152}$Eu decay rates show a temporary increase over the same time interval, and these two \textsc{roi}s arise from the daughter of the EC decay mode. The third \textsc{roi}, corresponding to the $\beta^-$ decay mode of $^{152}$Eu, does not exhibit such behavior. It should be emphasized that all $^{152}$Eu \textsc{roi}s are from the same detector with electronically-locked gain. \\

In any case, these fluctuations prevent meaningful half-life determinations in the early reactor-\textsc{off} period. Note that the statistical sensitivity of our detectors makes part-per-thousand changes readily visible. The coincidence of the counting rate fluctuations with the solar flare is intriguing in light of Refs. \cite{24,25}, but at this point is assumed to be due to the temperature excursion. This excursion after reactor shutdown may be responsible for the fluctuations in counting rate during that period. It is notable that the first \hbox{5 $\degree$C} temperature rise-and-plateau coincides with very little fluctuation in the $^{60}$Co counting rate, which is dominant during the second rise and peak period. This fact would appear to be inconsistent with a linear relationship between response change and temperature change. Among other lessons derived from the Phase I experiment was importance of proper temperature control, and this led to significant improvements in the design of our Phase II experiment described below.

\begin{table*}
\caption{Phase I Fits During the Reactor-\textsc{on} Period (3/23 -- 3/25)}
\label{tab:phase1fits}
\setlength{\extrarowheight}{3pt}
\begin{ruledtabular}
\begin{tabular}{crrrrr} 
 Isotope  & Feature (keV) & Half-life (\textsc{on})& $\pm$ stat. &  $\chi^2$/dof  & Published half-life\\
\hline
$^{60}$Co & $\beta^{-}$ 1173 & 5.28 a& 0.13& 1.13& 5.27 a\\
& 1333 & 4.87 a& 0.12& 1.17& \\
 \hline
 
$^{152}$Eu & $\beta^{+}$ EC \ \ 46 & 10.95 a& 0.40& 1.10& 13.54 a\\
& 122 & 11.22 a& 0.56& 1.01& \\
 &$\beta^{-}$ 344 & 10.23 a& 0.57& 1.08& \\
 \hline
$^{54}$Mn & EC 835 & 287.2 d& 0.60 & 1.46& 312 d\\
\end{tabular}
\end{ruledtabular}
\end{table*}

\begin{table*}
\centering
\caption{Detector Layout}
\label{tab:detector-layout}
\setlength{\extrarowheight}{5pt}
\begin{tabular}{|l|l|l|l|}
\hline
Detector 5 $^{241}$Am  & Detector 6 $^{22}$Na & Detector 7 $^{54}$Mn & Detector 8 Background \\ \hline
Detector 1 Background & Detector 2 $^{54}$Mn & Detector 5 $^{60}$Co & Detector 5 $^{152}$Eu     \\ \hline
\end{tabular}
\end{table*}

Notwithstanding the advantages of working with $^{54}$Mn, its relatively short half-life compared to the duration of the experiment revealed the rate-dependent nonlinearity discussed earlier.  The effect became clear immediately after our first \textsc{on} run, which extended from 15-20 March 2014. As can be seen from Table \ref{tab:phase1fits}, our determined $^{54}$Mn half-life was (287.2 $\pm$ 0.3) d which was significantly shorter than 312 d. Subsequent half-life measurements during both reactor \textsc{on} and reactor \textsc{off} periods yielded $^{54}$Mn half-life determinations which were consistently shorter than 312 d, but which approached the published value more closely as the source decayed.  The analysis procedure to correct for the observed detector-induced non-exponential counting rates is discussed in Section \ref{sec:AlphaDiscussion}.\\


\section{Phase II Experiment}

Phase II of our experiment, from which our results are derived, took place between August 2014 and March 2015 at \textsc{hfir}.  Analysis of $^{152}$Eu data was omitted due to the relatively low energy of the chosen peaks, where the background is greater.  This experiment included eight of the previously described 2 inch NaI(Tl) scintillation detectors with digi\textsc{base}s.  Each detector---except for the two background detectors---had a radioactive source fixed securely to the front of the detector as noted above. For Phase II, to accomodate the upcoming \textsc{prospect} neutrino experiment,  the detectors were moved from site EF-4 to site EF-3 of the \textsc{hfir} reactor, with the distance  from the reactor core increased from \hbox{5.8 m} to \hbox{6.6 m}, reducing the $\overline{\nu}_e$ flux to some 46 times the solar $\nu$ flux at Earth.  A new lead cave was constructed that consisted of two levels vertically with four detectors on each level.  A \hbox{2 inch} thick polystyrene insulating box was built around the lead cave and maintained at a stable temperature of \hbox{$20.00\pm 0.03$ $\degree$C} using a \textsc{teca} thermo-electric unit with Watlow PID controller.  A tabular representation of the detectors' set up is shown in Table \ref{tab:detector-layout}.\\ 

From differential thermal expansion of the NaI-photube assembly relative to the enclosing aluminum can, we estimate motion of the source relative to the NaI to be 1.95$\times10^{-3}$ mm/K. At 4$\%$ per mm of source motion, the resulting fractional change in solid angle subtended by the NaI is -7.8$\times10^{-5}$/K. Thermal swelling of the NaI increases the solid angle by 9.5$\times10^{-5}$/K. The resulting net fractional change in counting rate is 1.7$\times10^{-5}$/K, or 5.1$\times10^{-7}$ per 0.3 K, which is negligible.\\

Each of the eight detectors was connected to one of the four PCs at EF-3 via USB, and each was programmed to run for one hour live-time intervals. As noted above, live-time is the real-time as measured in the laboratory minus the dead-time as reported by the digi\textsc{base}. Each detector thus runs for an interval greater than one hour, where the total real time is equal to the fixed one hour of live-time plus the variable amount of dead-time added by the detector. During this interval, the detector saves the output from the Multi-Channel Analyzer into a time integrated 1024-bin histogram of energy. At the end of the time interval, information from this histogram is saved into two different files: one report file and one spectrum file. The spectrum file is simply a list of each of the 1024 bins and the counts recorded in each of those bins. The report file gives a shorter summary of the data (see below). In both files, basic information about the time interval is recorded, such as the date and time at which the measurement interval began, as well as both the real-time and live-time for the interval.\\

Along with the detector hardware, \textsc{ortec} also provides the \textsc{maestro} software in order to conveniently interact with their detectors. This software allows the user to set the high voltage, gain, gain locking, and other properties of each detector. In addition, during a measurement interval, the energy histogram (spectrum) is displayed on the screen optionally in either a linear or log scale. In order to make data acquisition and analysis easier, \textsc{maestro} allows the user to set custom \textsc{roi}s on the spectra for each detector.  At the end of each time interval when the report files are published, the total counts for each \textsc{roi} are summed over the specified energy range and reported. In addition to the integrated counts, \textsc{maestro} attempts to fit a peak for the given \textsc{roi} when a peak can be identified. \textsc{Maestro} reports a best fit for the centroid, full width at half max, and full width fifth max, of the peak.\\

Focusing on the \textsc{roi}s allows the gain locking and zero locking algorithms provided in the {\textsc{ortec}} digi\textsc{base}s to be implemented. The gain locking and zero locking algorithms prompt the user to identify an \textsc{roi} containing a peak to which the software will then  lock. As a counting interval begins, the gain locking software continuously attempts to find a peak within the locked \textsc{roi} (the user also sets the width, in bins, of the peak fitting region). If the \textsc{roi} contains a peak, the algorithm will adjust the fine gain settings of the detector in order to align the measured peak center with the peak center set by the user when the gain was originally locked.  In high statistics running with gain locking, the fitted peak location is stable to a small fraction of one bin.  Zero locking ensures that the zero of the detector does not drift---if an identifiable peak exists. The combination of gain locking and zero locking makes it possible to essentially eliminate drifts in the gain and zero of the detectors, making measurements much more accurate in the long run.\\


\section{Correcting for Detector-induced Rate-dependent Distortions in the Measured Counting Rate}
\label{sec:AlphaDiscussion}

As mentioned earlier, pile-up and dead time are known problems in counting-detectors, which worsen with higher counting rates.  The digi\textsc{base} corrects for the dead time, using a variant of the Gedcke-Hale algorithm.  The \textsc{Maestro}/digi\textsc{base} system cannot correct for pileup effects.  Pileup occurs when two photons which hit the detector close in time are not resolved, and the sum of their energies is registered as one count at a higher energy, and the other count is lost.  The pileup effect is a rate-dependent convolution of the energy spectrum with itself, and whether the net pileup adds to, or subtracts from, a given \textsc{roi} is a complicated issue.  In all of the \textsc{roi}s used in our present data,  pileup appears to somewhat mimic small excess deadtime corrections of less than 2$\mu$s per count.\\ 

To achieve sensitivities better than one part in 10$^4$, we have chosen rather high overall counting rates, up to some 35 kHz, thereby entailing up to 13$\%$ dead time and significant pile-up.  When counting a given radioactive sample for a significant fraction of one half-life, we observe a distortion relative to a pure exponential decay curve.    The counting curve is steeper than exponential early in the measurement, and less steep than exponential late in the measurement.  After detrending the data using a pure exponential, the residuals then show a characteristic U shape.\\

This distortion manifests itself in our data through half-life measurements that are systematically low (larger decay constants), and which steadily approach the published half-life values as the sample decays (as observed in our Phase I experiment).  These changes in decay constant are obviously purely instrumental, and have nothing to do with putative variations in the actual decay constant.  The distortion can be well modeled by a correction factor 1/(1+$\alpha'{dN}/{dt}$) to be applied to the data to bring them into a pure exponential  form.  The parameter $\alpha'$ is typically 1.2 to 1.8$\mu$s where ${dN}/{dt}$ is the overall counting rate into the electronics.  This pile-up correction factor is algebraically equivalent to an excess added deadtime of  1.2 to 1.8$\mu$s to the run time for each count, in addition to approximately 4 $\mu$s of actual deadtime correctly assigned by the detector system. We emphasize that this form is also a standard parameterization of the effects of pile-up. As will be seen below, this form, or its equivalent (to a good approximation) form exp($\alpha'{dN}/{dt}$) adequately removes the observed rate-dependent distortions.\\


\section{Analysis Procedures}

If the decay parameter, $\lambda$, changes in the presence of the reactor antineutrino flux, we expect different slopes of the decay curve between reactor-\textsc{on} and \textsc{off} periods.  Note that we measure the counting rate, which (if the solid angle subtended by the detector does not change) is proportional to ${dN}/{dt}$  where $N(t)$ is the number of source nuclei as a function of time.  
If $\lambda$ is stepwise constant, i.e. alternates between values $\lambda$ and $\lambda'$ for reactor \textsc{off} and \textsc{on}, respectively, then the counting rate is proportional to 

\begin{equation} \label{eq:analysis-1}
\dot{N(t)}=\frac{dN}{dt}=\frac{d}{dt}(N_0 e^{-\lambda t})= -\lambda N_0 e^{-\lambda t}\ \  (\textsc{off}),
\end{equation} 

At the point when the reactor turns \textsc{on} if $\lambda$ changes to $\lambda'$, $N(t)$ does not change instantaneously, hence the step in $\lambda$ must cause a step in the counting rate, and we would expect there to be a step in the counting rate each time there is a transition between \textsc{on} and \textsc{off} or vice versa.  In practice, this is a much more sensitive way of searching for changes in $\lambda$ than through analysis of the slopes $dN/dt$ themselves over time.\\ 

To capture the entire set of effects with optimal use of the data, we perform a four-parameter “Global Fit” to the entire sequence of five \textsc{on} and four \textsc{off} reactor periods. The function used is a stepwise sequence of exponential decays with alternating decay parameters $\lambda$ and $\lambda(1 + \epsilon)$, where $\epsilon \equiv (\lambda_{\textsc{on}} -  \lambda_{\textsc{off}})/\lambda_{\textsc{off}}$, in an obvious notation.  The other two fitting parameters are $C_0$, the initial counts in one hour; and the distortion parameter $\alpha$ discussed in Section \ref{sec:AlphaDiscussion}.  The exact formulas used are given in Appendix \ref{app:GlobalFitting}.\\

The counting rate distortion factor used in the stepwise global formula is $e^{\alpha  C_{tot}(t)}$, where $C_{tot}(t) \equiv \dot{N}_{tot}(t)*3600s$ is the total number of counts in one hour; for clarity we have ignored counting inefficiencies. To a good approximation, the factor $e^{\alpha  C_{tot}}$ is identical to $(1+ \alpha' \dot{N}_{tot})$ (the exponent is typically 3$\%$ or less). $\alpha'$ and $\alpha$ differ by a factor of 3600 due to converting from counts per one-hour interval to Hz.  Either the $\alpha$ or the $\alpha'$ version can of course be used in the exponential form.  We choose the  $\alpha$ version for convenience.


\section{The Data}

\begin{figure*}
	\subfloat[]{\includegraphics[width=0.48\linewidth]{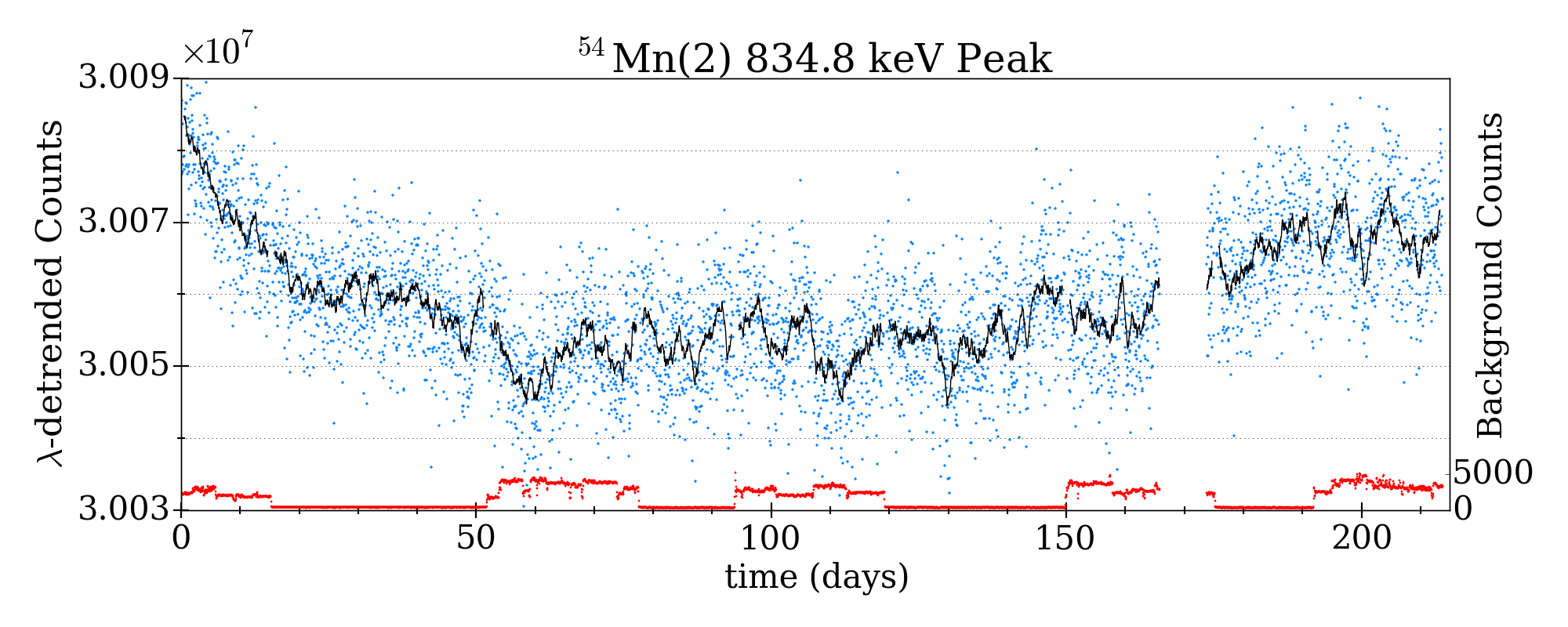}}
    \subfloat[]{\includegraphics[width=0.48\linewidth]{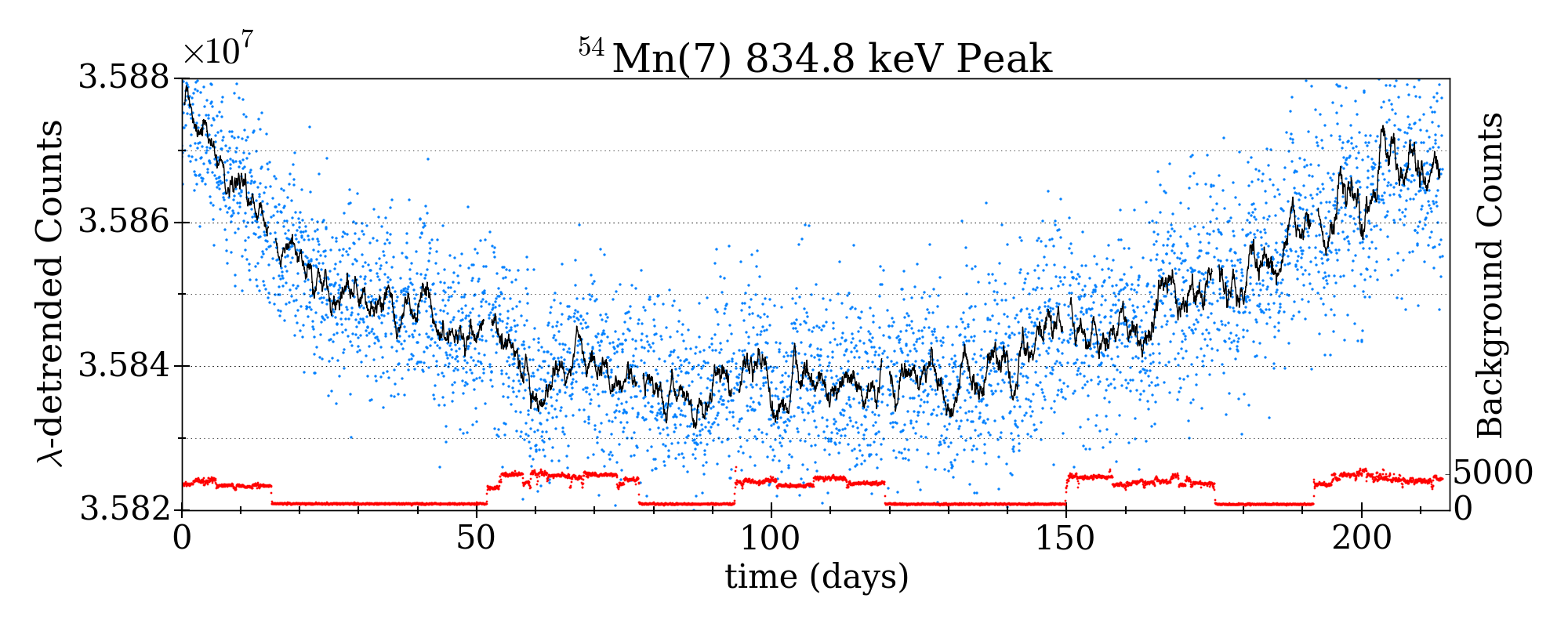}}
    
    \subfloat[]{\includegraphics[width=0.48\linewidth]{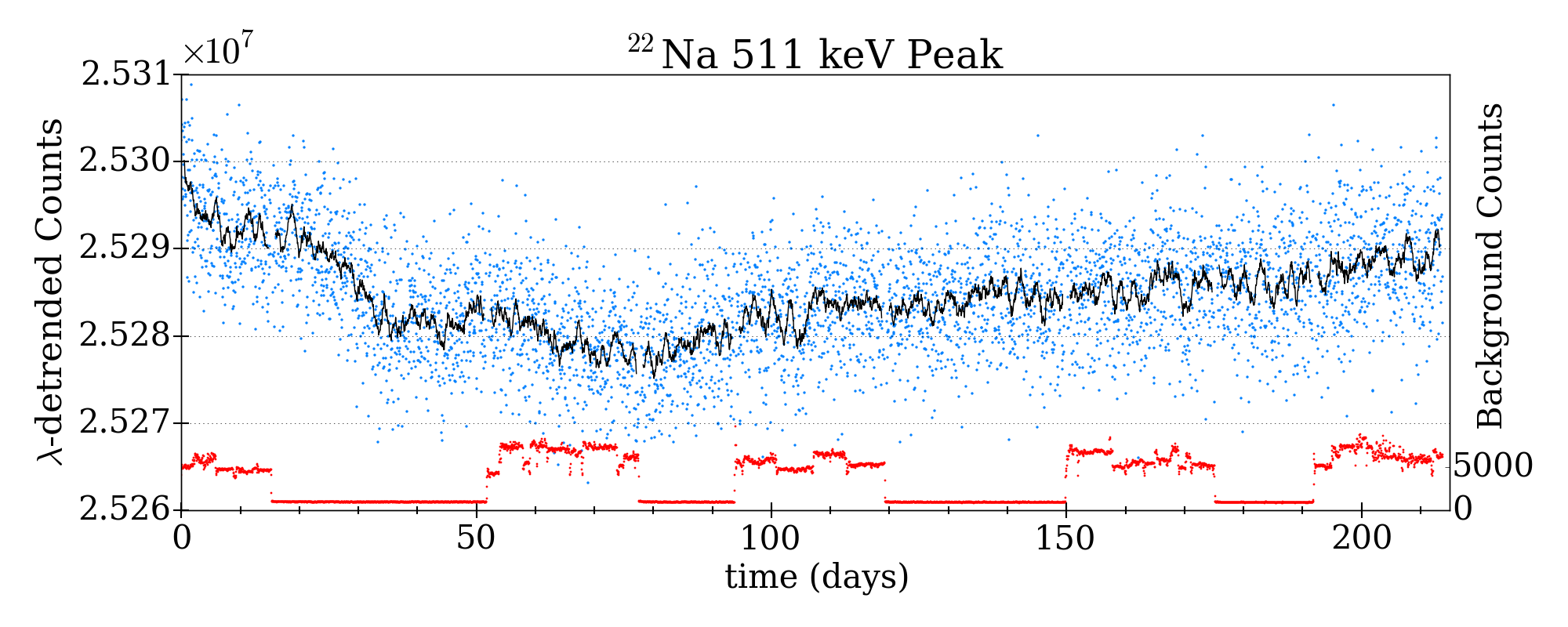}}
    \subfloat[]{\includegraphics[width=0.48\linewidth]{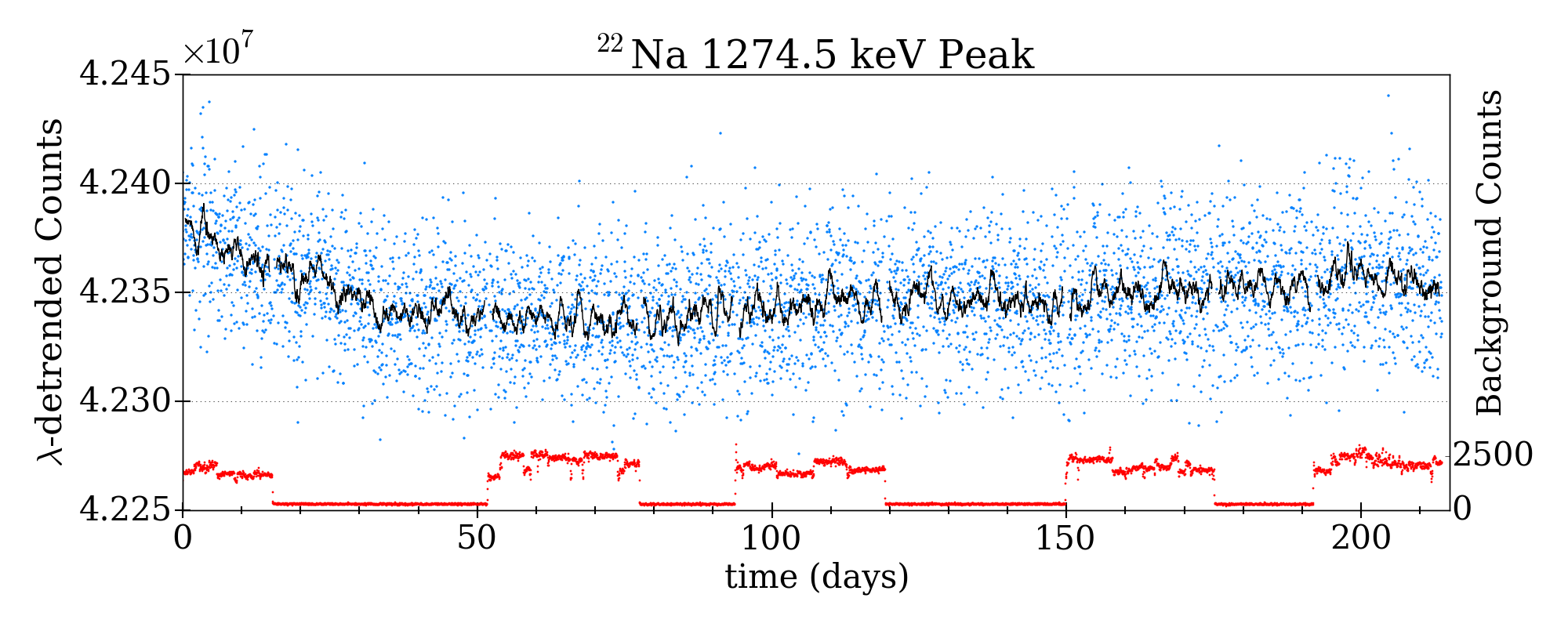}}
    
    \subfloat[]{\includegraphics[width=0.48\linewidth]{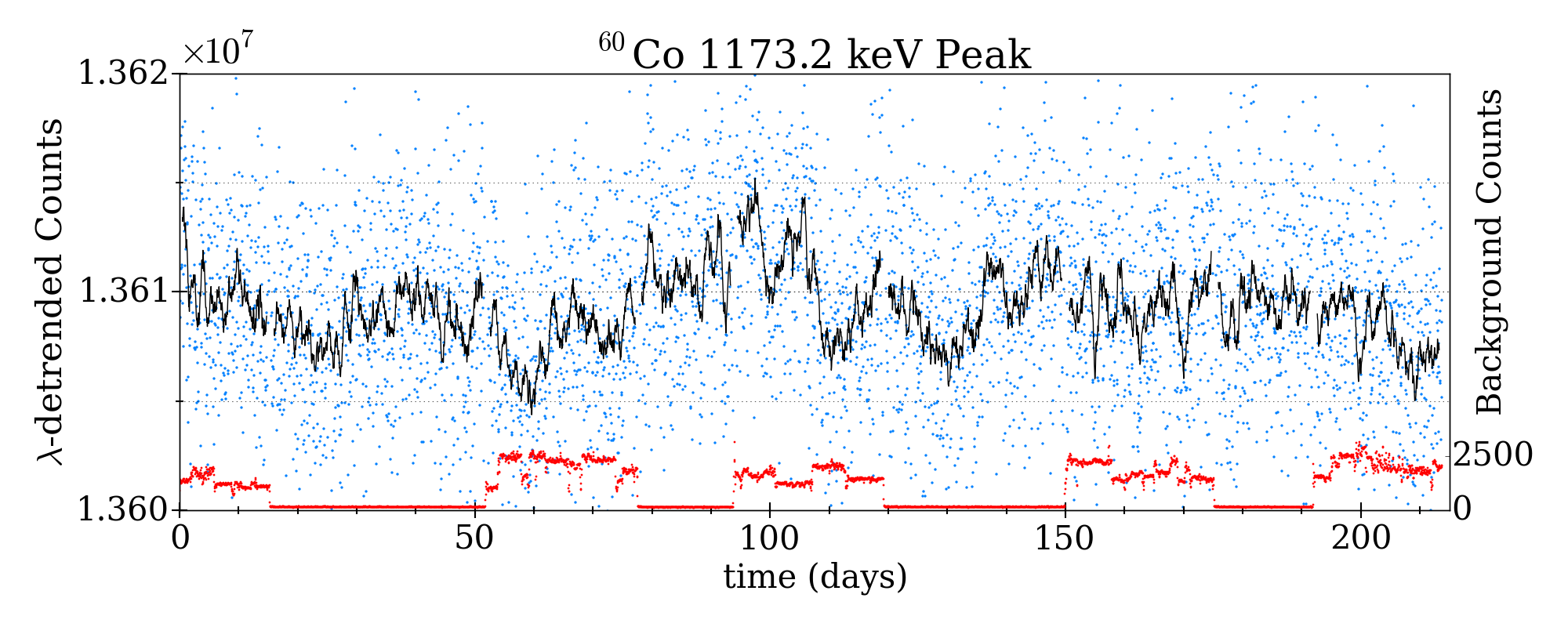}}
    \subfloat[]{\includegraphics[width=0.48\linewidth]{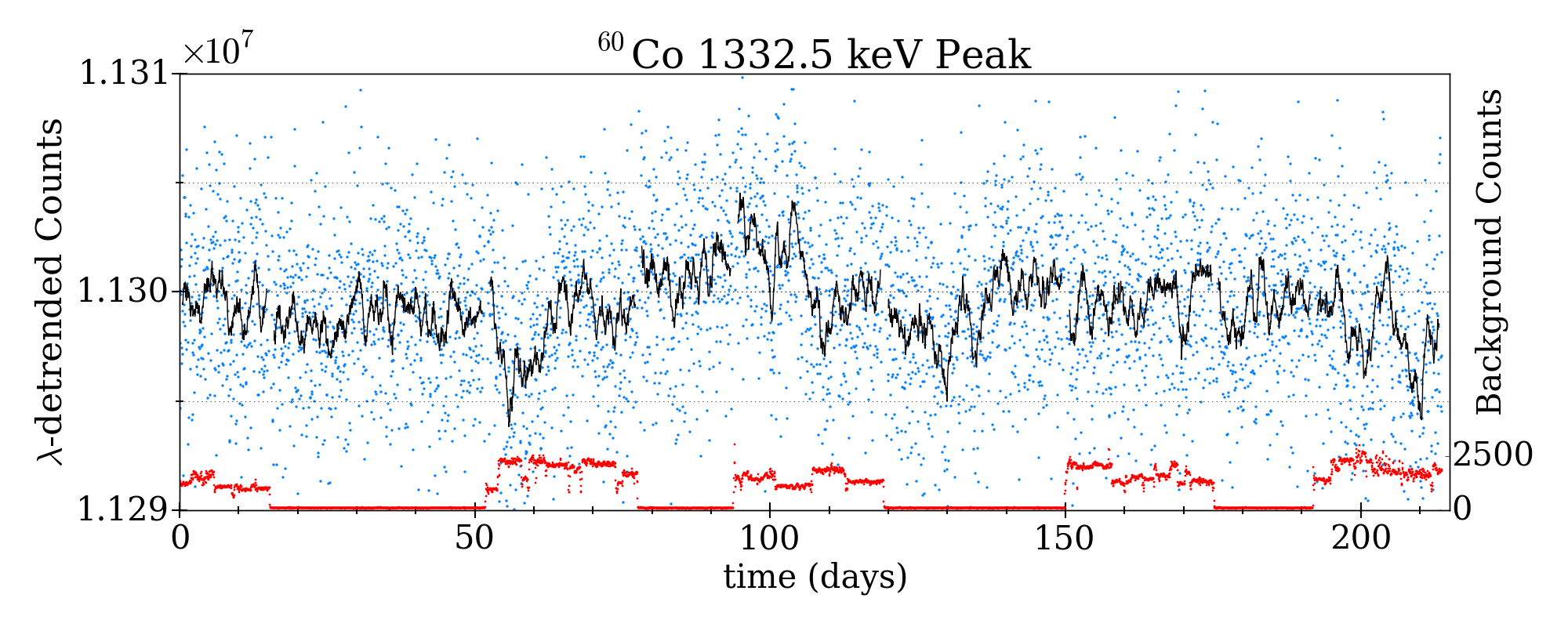}}
    \caption{Exponentially detrended hourly counts vs. time, [and the full vertical fractional intervals for]: (a) $^{54}$Mn Det.2 [2.0$\times$10$^{-3}$] (b) $^{54}$Mn Det.7 [1.7$\times$10$^{-3}$] (c) $^{22}$ Na annih. [2.0$\times$10$^{-3}$] (d) $^{22}$Na gamma [4.7$\times$10$^{-3}$] (e) $^{60}$Co low [1.4$\times$10$^{-3}$] (f) $^{60}$Co high [1.8$\times$10$^{-3}$].  The backgrounds which have been subtracted are shown in red, at an identical vertical scale. The heavy black line is a 20-point moving average.}	\label{fig:Detrended}
    \centering
\end{figure*}

In Figure \ref{fig:Detrended} we show the data from both $^{54}$Mn detectors, along with the $^{22}$Na and $^{60}$Co detectors (two peaks each) detrended by pure exponentials using the fitted values of $\lambda$.  These are simple 2-parameter fits (\hbox{$\alpha$ = 0} and \hbox{$\epsilon$ = 0}). The black curves are $20$-point moving averages. The residual U-shaped distortions are seen in the two $^{54}$Mn plots.  Given the longer half-lives of $^{22}$Na and $^{60}$Co, we expect such distortions to be less apparent since those counting rates have changed less during the experiment. \textsc{roi} backgrounds based on Detector 1 or 8 measurements are subtracted on an hourly basis.  The subtracted amounts are plotted below the de-trended source plots, with identical vertical scales for comparison. Modest uncertainties in the background subtractions will be dealt with in our treatment of systematic uncertainties, below. The full background treatment is rather involved, and is given in Appendix \ref{app:BackgroundSubtraction}.\\

\begin{figure*}
	\subfloat[]{\includegraphics[width=0.47\linewidth]{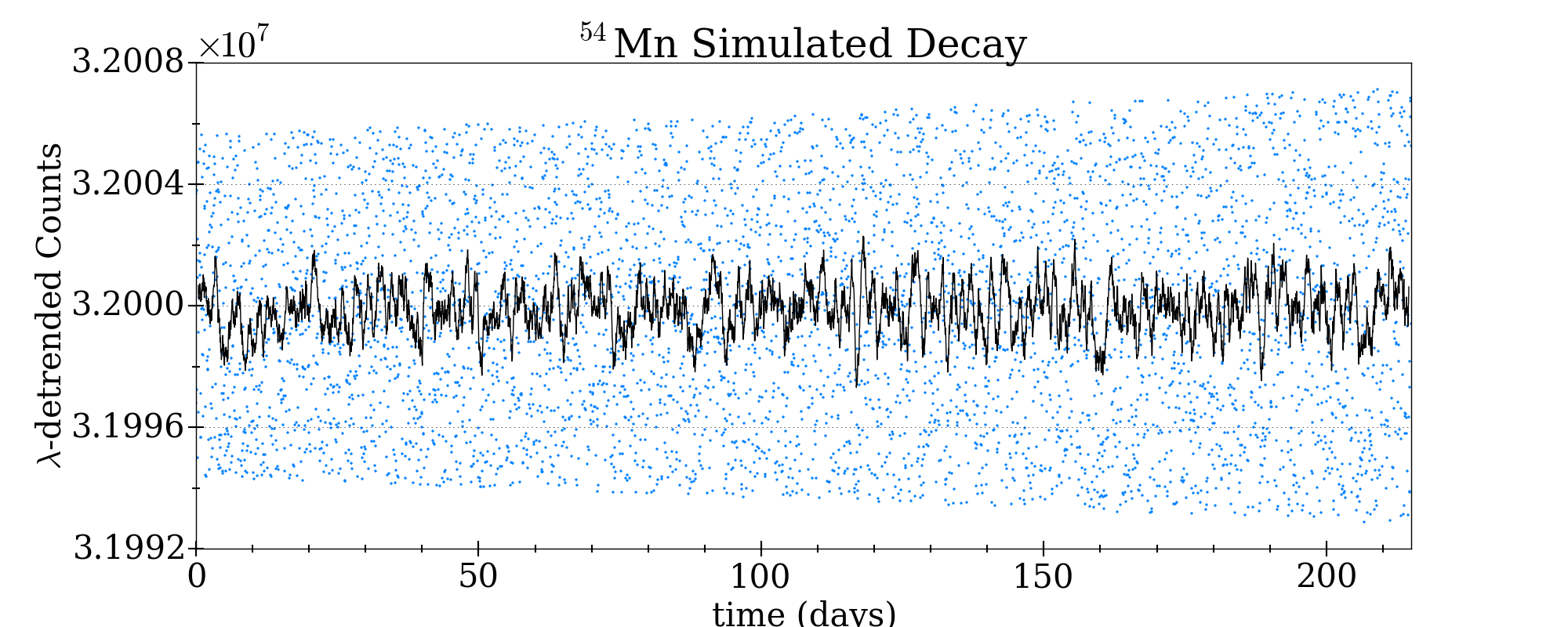}}
    \subfloat[]{\includegraphics[width=0.47\linewidth]{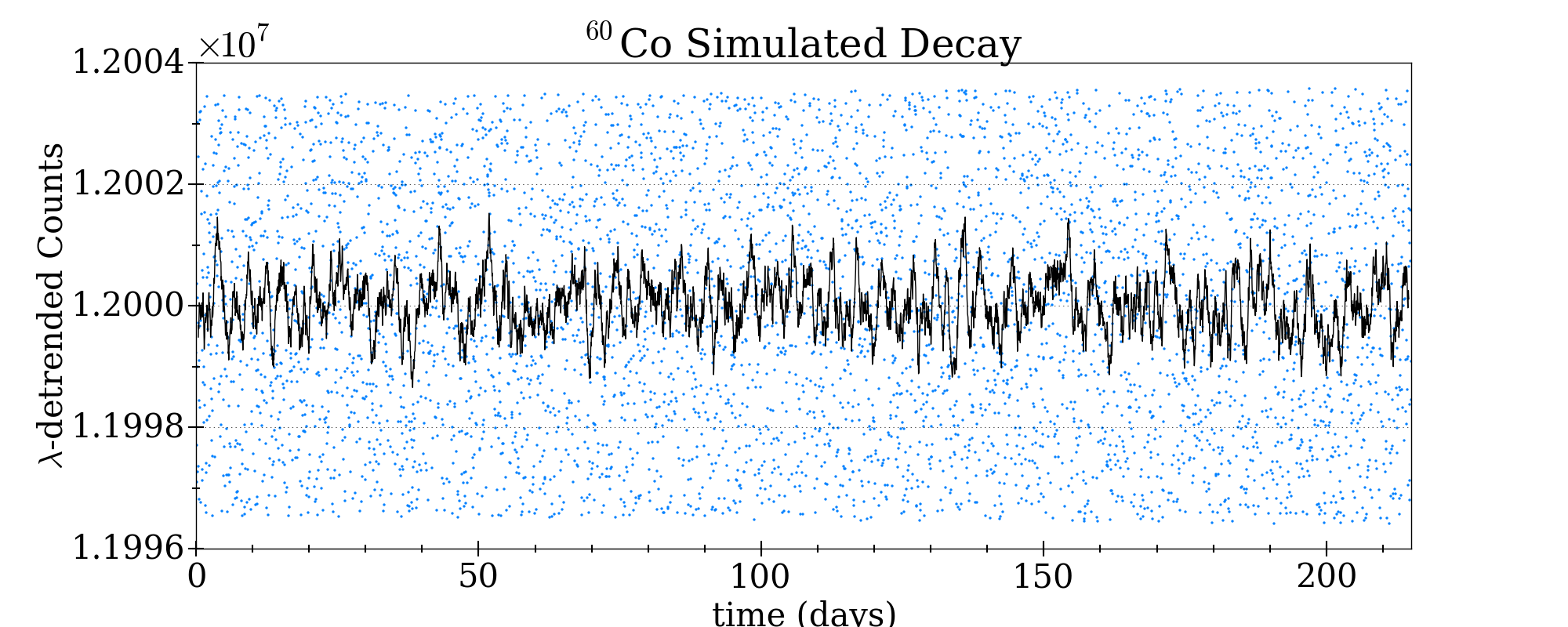}}
    
    \subfloat[]{\includegraphics[width=0.47\linewidth]{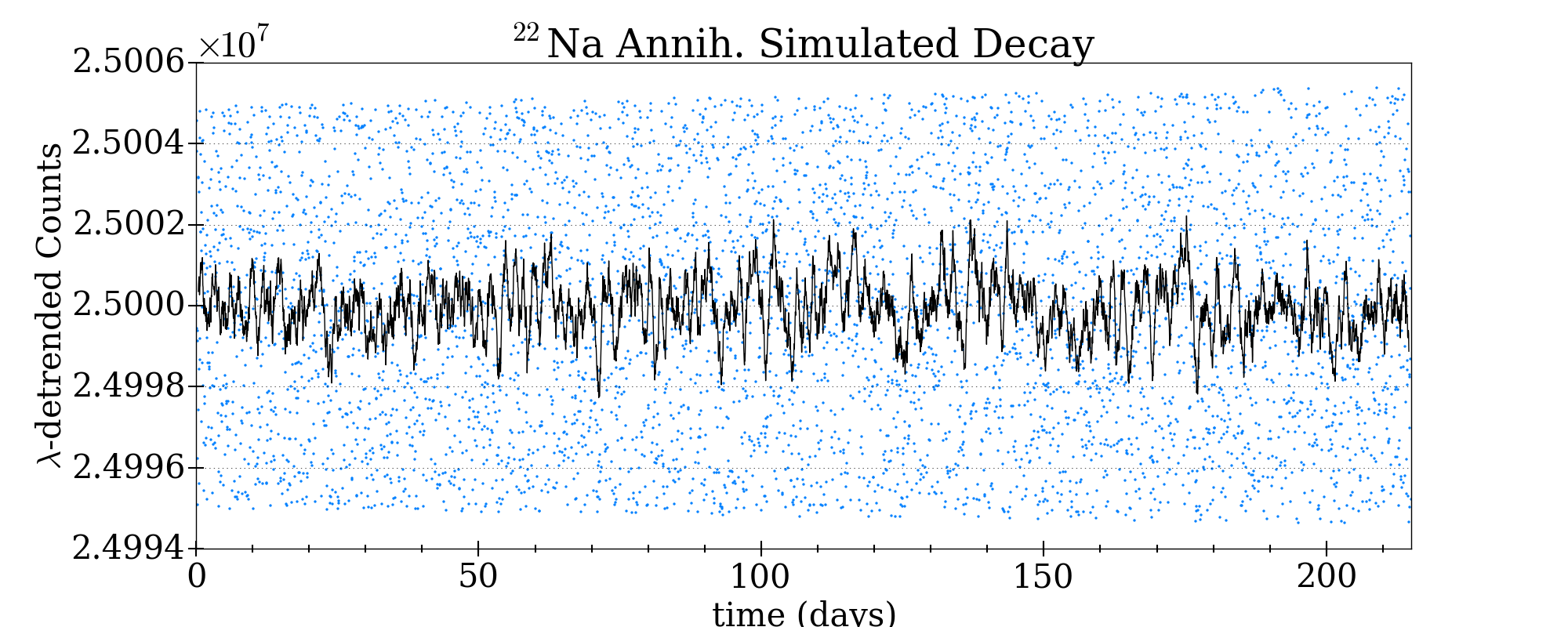}}
    \subfloat[]{\includegraphics[width=0.47\linewidth]{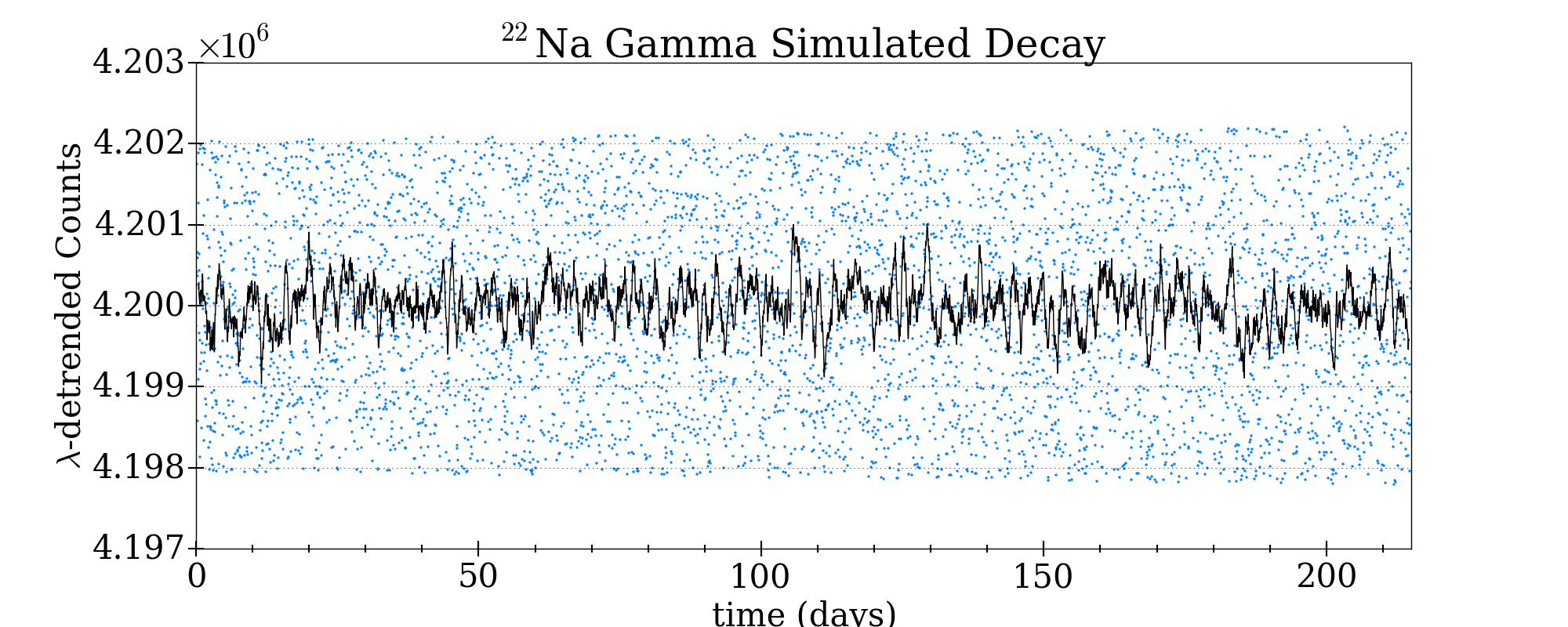}}
     \caption{Exponentially detrended Monte Carlo data series reflecting the Gaussian statistics of (a) $^{54}$Mn [3$\times 10^7$ hourly counts] (b) $^{60}$Co [1.2$\times 10^7$ hourly counts] (c) $^{22}$Na annih. [4.2$\times 10^6$ hourly counts] (d) $^{22}$Na gamma [2.5$\times 10^7$ hourly counts]. The total vertical fractional intervals are $4.3\times10^{-4}$, $6.7\times10^{-4}$, $4.8\times10^{-4}$, and $1.4\times10^{-3}$, respectively. The heavy black line is a 20-point moving average. }	\label{fig:MC}
    \centering
\end{figure*}

\begin{figure*}
	\subfloat[]{\includegraphics[width=0.47\linewidth]{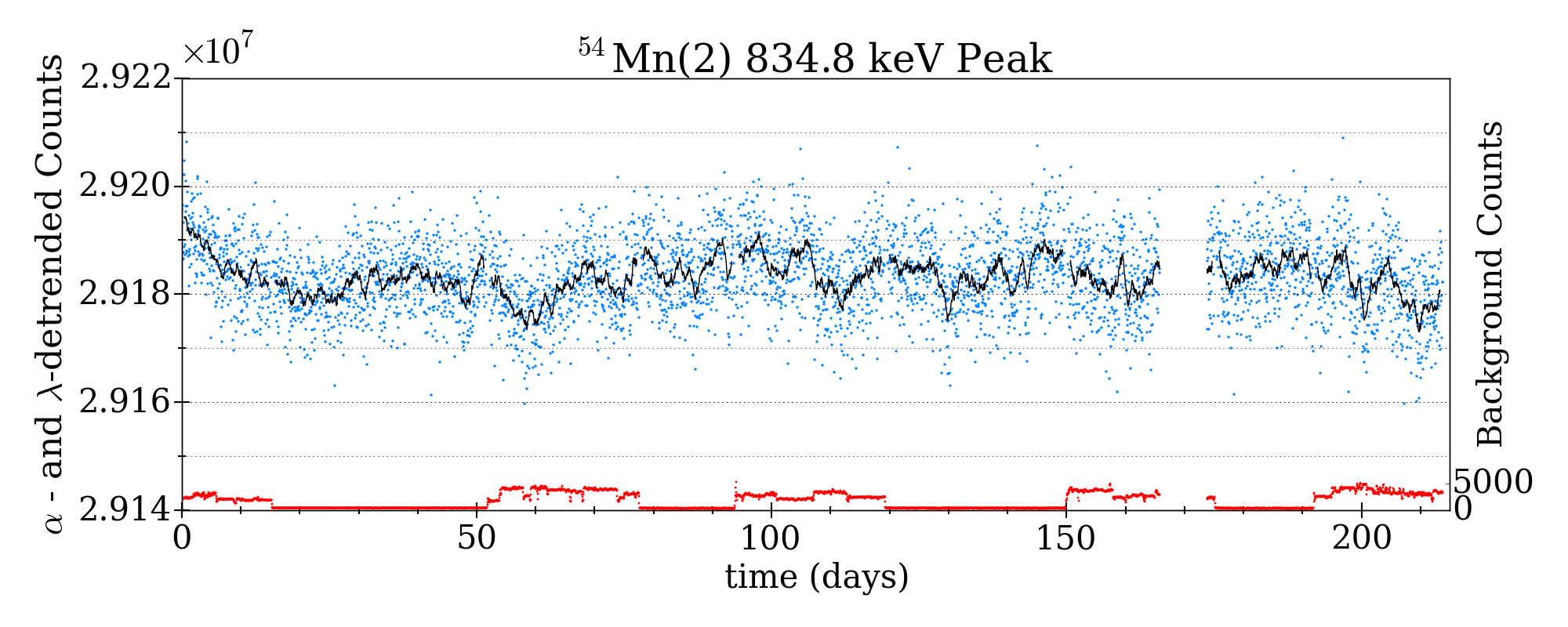}}        
    \subfloat[]{\includegraphics[width=0.47\linewidth]{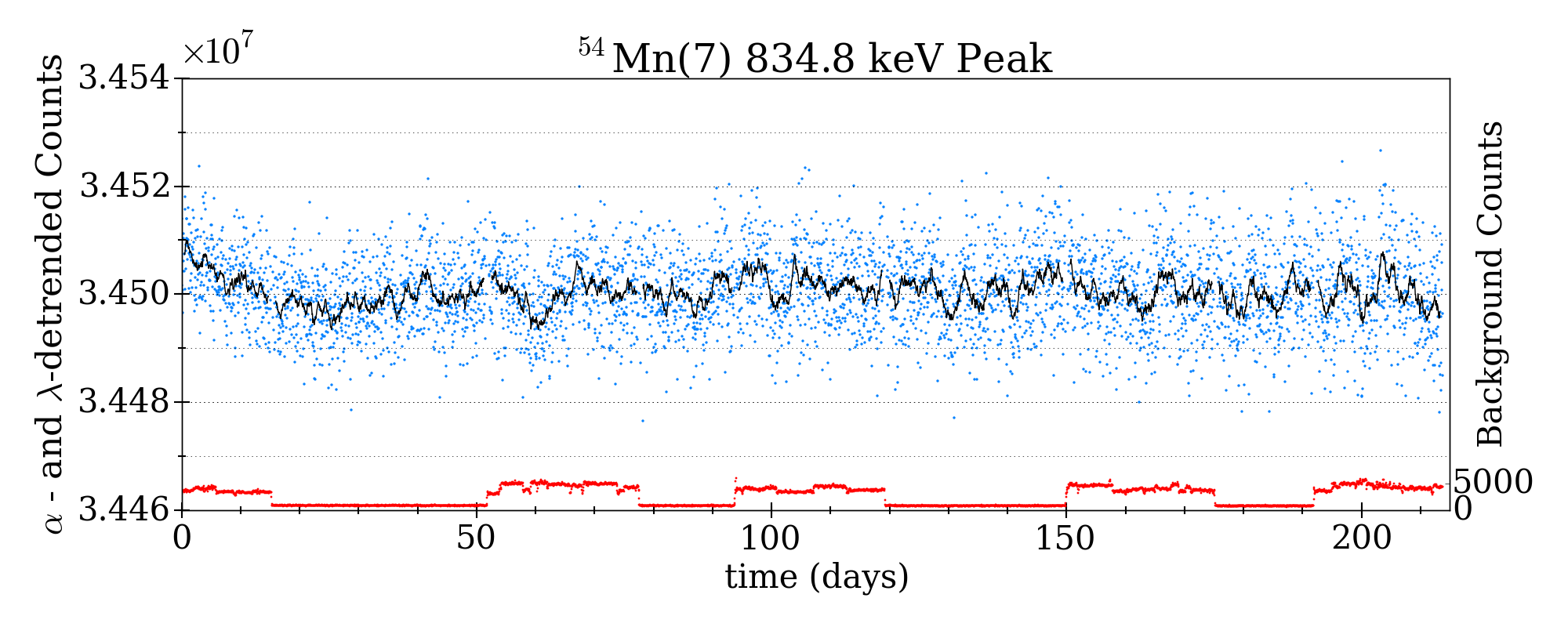}} 
    
    \subfloat[]{\includegraphics[width=0.47\linewidth]{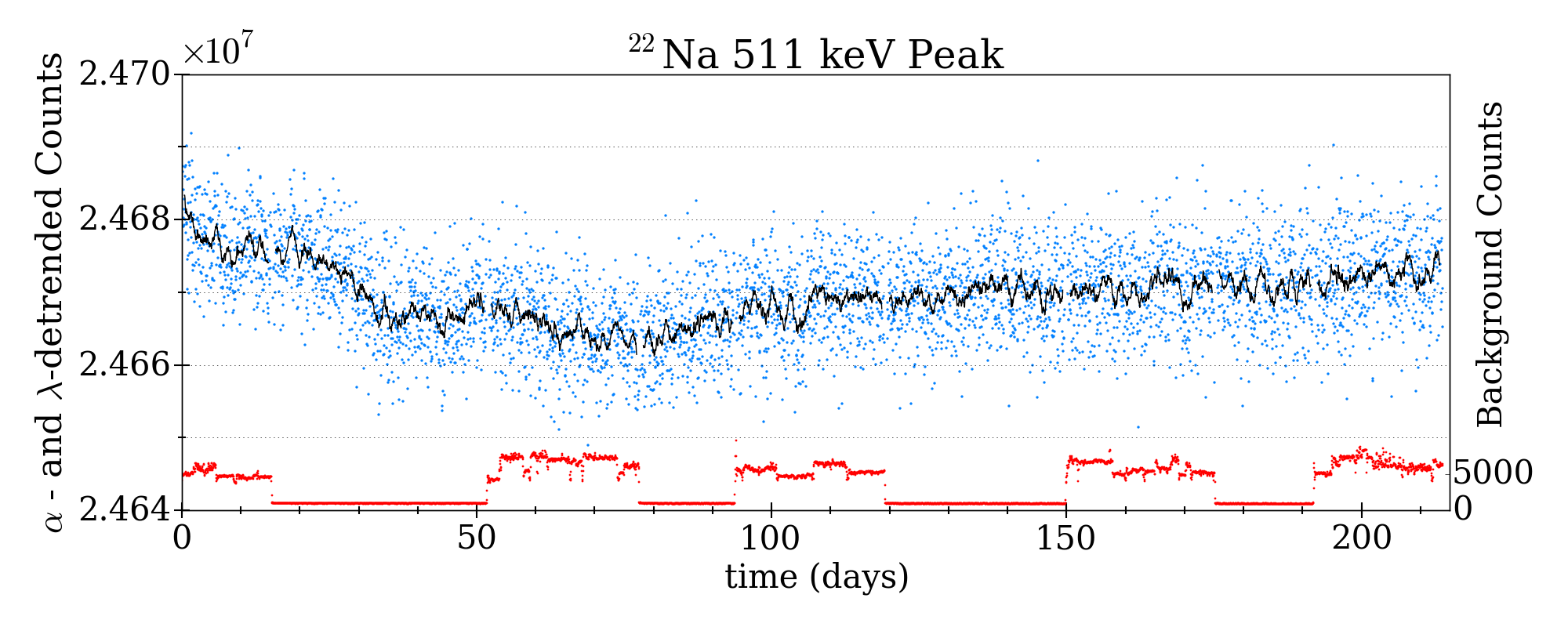}}    
    \subfloat[]{\includegraphics[width=0.47\linewidth]{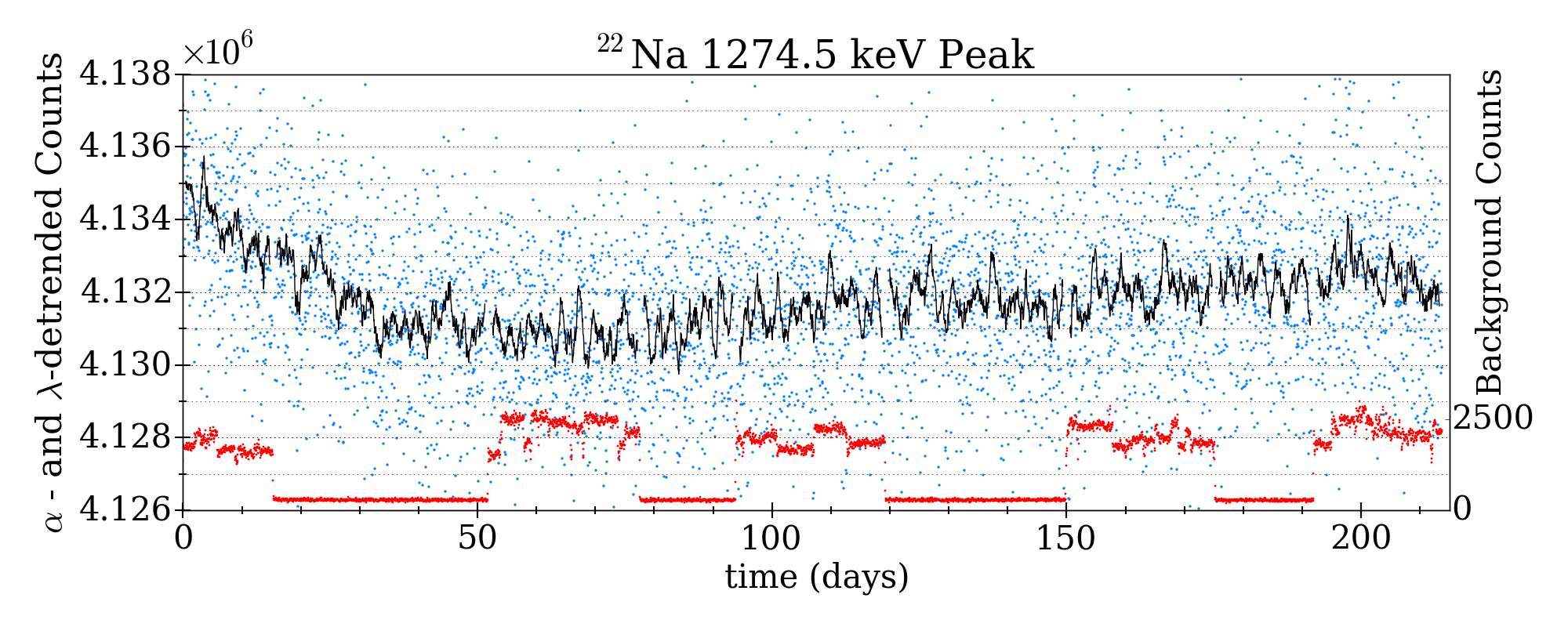}} 
    
    \subfloat[]{\includegraphics[width=0.47\linewidth]{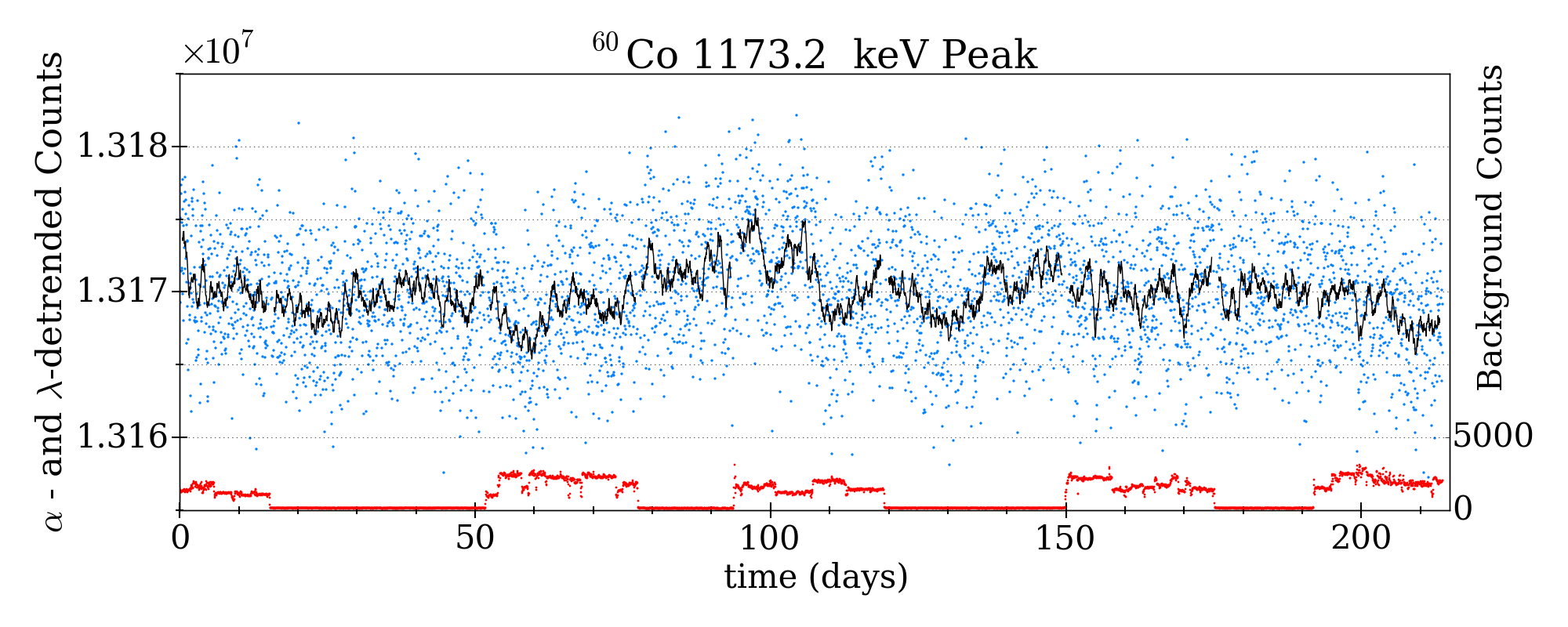}}    
    \subfloat[]{\includegraphics[width=0.47\linewidth]{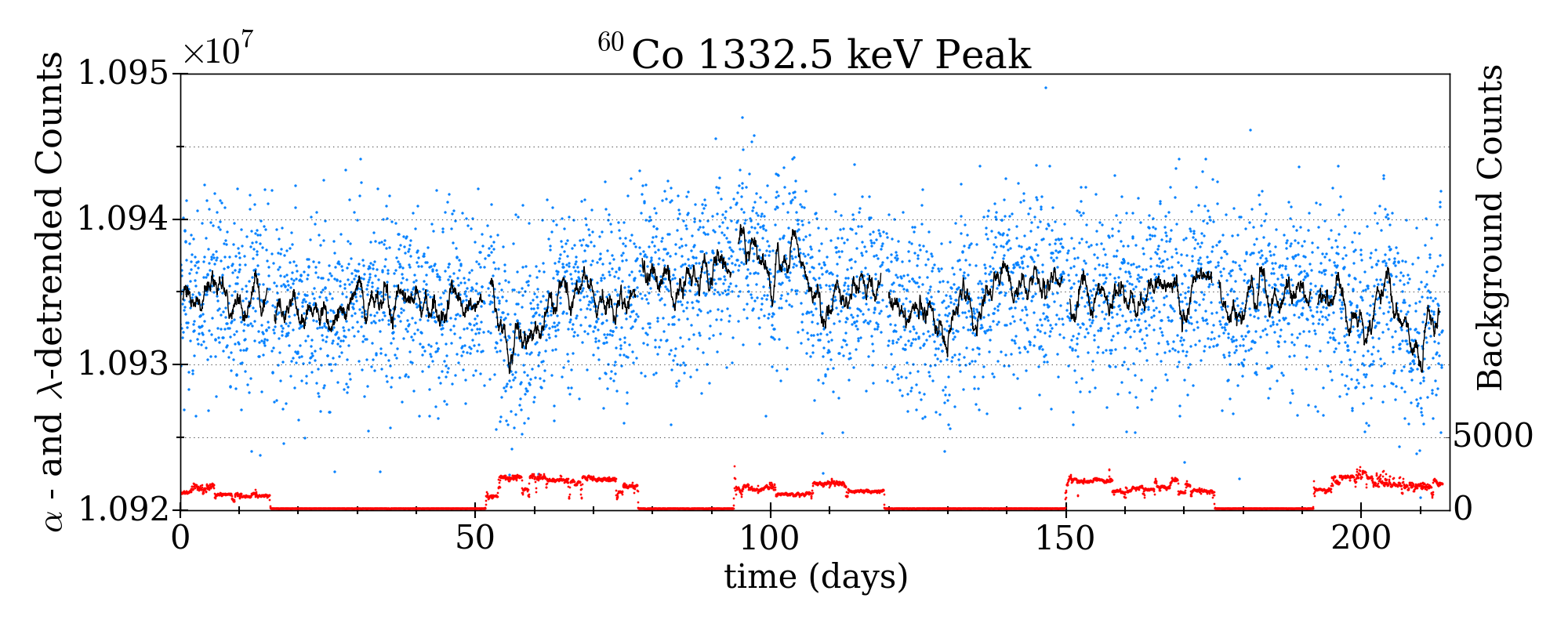}}
    
    \caption{Exponentially- and $\alpha$-detrended hourly counts vs. time, [and the full vertical fractional intervals for]: (a) $^{54}$Mn Det.2 [2.7$\times$10$^{-3}$] (b) $^{54}$Mn Det.7 [2.3$\times$10$^{-3}$] (c) $^{22}$ Na annih. [2.4$\times$10$^{-3}$] (d) $^{22}$Na gamma [2.9$\times$10$^{-3}$] (e) $^{60}$Co low [2.3$\times$10$^{-3}$] (f) $^{60}$Co high [2.7$\times$10$^{-3}$].  The backgrounds which have been subtracted are shown in red, at an identical vertical scale. The heavy black line is a 20-point moving average.}	\label{fig:AlphaDetrended}
    
\end{figure*}

Generally, the moving averages wander more than expected from purely statistical fluctuations.  To calibrate the eye, Figure \ref{fig:MC} shows Monte Carlo simulated pure exponential decays for the three isotopes, with Gaussian statistical fluctuations based on the hourly counts: 3$\times10^7$ ($^{54}$Mn); 1.2$\times10^7$ ($^{60}$Co); 2.5$\times10^7$ and 4.0$\times10^6$ for the $^{22}$Na annihilation and gamma peaks, respectively. The scales of the fluctuations are much smaller than those of the actual data. Even after de-trending the $\alpha$ distortion, as seen in Figure \ref{fig:AlphaDetrended}, these irregularities in the actual data are well above purely statistical, must be systematic, and are of unknown origin.  These systematics will be quantified below, to some extent, in three different ways using the data.  \\

The results of the Global Fits are shown in Table \ref{tab:GlobalFits}.  The stated errors are purely statistical, and the $\chi^2$/dof are all somewhat larger than, but reasonably close to,  1. The two main $^{60}$Co peaks are fitted separately, as are the $^{22}$Na positron annihilation peak and the gamma peak from the daughter nuclide.  The values of $\epsilon$ for the two $^{22}$Na peaks are in good statistical agreement with each other, as are those for the two $^{60}$Co peaks, as shown in the $\Delta\epsilon$ column of Table \ref{tab:allepsilons}.  There is a high correlation between $\lambda$ and $\alpha$.  For the two longer-lived isotopes, which are relatively insensitive to $\alpha$, the best full fits give unreasonable values for both parameters. Hence, for these fit results, we fix the values of $\alpha$ to be 3.35$\times$10$^{-10}$ ($\alpha'=1.2$ $\mu s$) as determined with good agreement by the fits for the two $^{54}$Mn detectors. With this choice, the half-lives of all four sources are fairly close to the published values, but differ from them (and in some cases, from each other) by many statistical standard deviations. \\

If we fix the decay constants to the published values, as shown in Table \ref{tab:GlobalFitsNew},  the values of $\chi^2$/dof increase by less than 1$\%$  for Co and Na, and by 3.6$\%$ and 6.7$\%$ for the two Mn detectors.  The values of $\alpha$ change modestly, but remain in the range of (3.3 to 5.5)$\times$10$^{-10}$.  The values of $\epsilon$ change, and  we take the changes in $\epsilon$ to be one of three estimates of the unknown systematic errors: (1) "Fit Change" .  The other two estimates of systematics are (2) from the results of (2) a "Shuffle Test" to be described below, and (3) from the difference  $\Delta\epsilon$ for each isotope; however,  $\Delta\epsilon$ for Na and Co are not statistically significant, hence unlikely to represent systematic errors, and so to avoid "double counting" in these two cases, we propagate the $\Delta\epsilon$ column in Table \ref{tab:allepsilons} only for the comparison of the two Mn detectors, where $\Delta\epsilon$ is six standard deviations and a clear indication of some (unknown) systematic effect(s) .     
\begin{table*}
\caption{Global Fit Results}
\label{tab:GlobalFits}
\setlength{\extrarowheight}{3pt}
\begin{ruledtabular}
\begin{tabular}{ccrrrrrrr}

 Isotope  & Feature & $\lambda$ (days$^{-1}$) & C$_0$ & $\alpha$ (s) & $\epsilon$ & $\chi^2$/dof & Half-life (d) & Published (d)\\
\hline
$^{54}$Mn & Det. 2 & 2.23175$\times10^{-3}$&  2.92$\times10^7$ & 3.224$\times10^{-10}$ & -1.21$\times10^{-5}$ & 1.295 & 310.58 & 312 \\

		  & $\pm\sigma_{stat}$\footnote{Statistical errors are given below fitted values in every row.} & \ 7.5$\times10^{-7}$ & 1.2$\times10^4$ & 4.4$\times10^{-12}$ & 6.2$\times10^{-6}$ &  & 0.07&\\

  		  & Det. 7 & 2.23392$\times10^{-3}$& 3.45$\times10^{7}$ & 3.472$\times10^{-10}$ & $4.05\times10^{-5}$ & 1.150 & 310.28 &\\

		  &  & 7.0$\times10^{-7}$ &  1.3$\times10^{4}$ & 3.4$\times10^{-12}$ & 5.7$\times10^{-6}$ &  & 0.07&\\
          \hline

$^{22}$Na & 511 keV & 7.29416$\times10^{-4}$\ & 2.47$\times10^7$ & 3.35$\times10^{-10}$ & 2.96$\times10^{-5}$ & 1.490 & 950.28& 949.7\\
&  & 4.9$\times10^{-8}$ & 1.6$\times10^2$ & Fixed & 6.1$\times10^{-6}$ &  & 0.04 &\\
& 1275 keV & 7.39604$\times10^{-4}$ & 4.13$\times10^6$ & 3.35$\times10^{-10}$ & 6.3$\times10^{-5}$ & 1.143 & 937.19&\\
&  & 1.2$\times10^{-7}$ & 6.5$\times10^1$ & Fixed & 1.5$\times10^{-5}$ &  & 0.11&\\
\hline

$^{60}$Co & 1173 keV & 3.66119$\times10^{-4}$ & 1.32$\times10^7$ & 3.35$\times10^{-10}$ & -1.26$\times10^{-5}$ & 1.166 & 1893.23&1924.9\\
&  & 6.7$\times10^{-8}$ & 1.2$\times10^2$ & Fixed & 8.3$\times10^{-6}$ &  & 0.24&\\
& 1333 keV & 3.67259$\times10^{-4}$ & 1.09$\times10^7$ & 3.35$\times10^{-10}$ & 1.74$\times10^{-7}$ & 1.116 & 1887.35&\\
&  & 7.4$\times10^{-8}$ & 1.1$\times10^2$ & Fixed & 9.3$\times10^{-6}$ &  & 0.26&\\

\end{tabular}
\end{ruledtabular}
\end{table*}

\begin{table*}
\caption{Global Fit Results With $\lambda$ Fixed to Published Value} 
\label{tab:GlobalFitsNew}
\setlength{\extrarowheight}{3pt}
\begin{ruledtabular}
\begin{tabular}{ccrrrrrc}

 Isotope  & Feature &  C$_0$ &  $\lambda$ (days$^{-1}$) &$\alpha$ (s) &$\epsilon$ & $\chi^2$/dof & Published Half-life\\
\hline
$^{54}$Mn & Det. 2 & 2.90$\times10^{7}$& 2.221$\times10^{-3}$ & 3.858$\times10^{-10}$ & -2.23$\times10^{-5}$ & 1.341 &  312 d \\

		  & $\pm\sigma_{stat}$\footnote{Statistical errors are given below fitted values in every row.} & 6.5$\times10^{2}$ &Fixed & 2.9$\times10^{-13}$ & 6.1$\times10^{-6}$ &  &   \\

  		  & Det. 7 & 3.43$\times10^{7}$& 2.221$\times10^{-3}$ &4.102$\times10^{-10}$ & 2.75$\times10^{-5}$ & 1.227 &  \\

		  &  & 6.9$\times10^{2}$ & Fixed  &2.2$\times10^{-13}$ & 5.6$\times10^{-6}$ &  &   \\
          \hline

$^{22}$Na & 511 keV & 2.47$\times10^{7}$\ & 3.601$\times10^{-4}$ & 3.265$\times10^{-10}$ & 3.05$\times10^{-5}$ & 1.494 & 2.60 a  \\
&  & 1.6$\times10^{3}$ & Fixed &9.6$\times10^{-13}$ &  6.1$\times10^{-6}$ & &  \\
& 1275 keV & 4.07$\times10^{6}$ & 3.601$\times10^{-4}$& 5.249$\times10^{-10}$ & 5.33$\times10^{-5}$ & 1.127  \\
&  & 6.5$\times10^{2}$ & Fixed &2.3$\times10^{-12}$ & 1.5$\times10^{-5}$ & &  \\
\hline

$^{60}$Co & 1173 keV & 1.30$\times10^{7}$ & 7.299$\times10^{-4}$ & 5.051$\times10^{-10}$ & -1.24$\times10^{-5}$ & 1.161 & 5.27 a \\
&  & 2.3$\times10^{3}$ & Fixed & 1.9$\times10^{-12}$ & 8.3$\times10^{-6}$ & &  \\
& 1333 keV & 1.07$\times10^{7}$ & 7.299$\times10^{-4}$& 5.371$\times10^{-10}$ & -6.82$\times10^{-7}$ & 1.112 &  \\
&  & 2.1$\times10^{3}$ & Fixed & 2.1$\times10^{-12}$ & 9.4$\times10^{-6}$ & & \\

\end{tabular}
\end{ruledtabular}
\end{table*}

  A known systematic effect is due to the small uncertainties in the relative background levels in adjacent bays of the cave (see Appendix B), and, like the statistical errors, turns out to be negligible in quadrature with the three unknown-systematics estimates as we shall see below .  \\

\begin{figure}
	\subfloat[]{\includegraphics[width=0.47\textwidth]{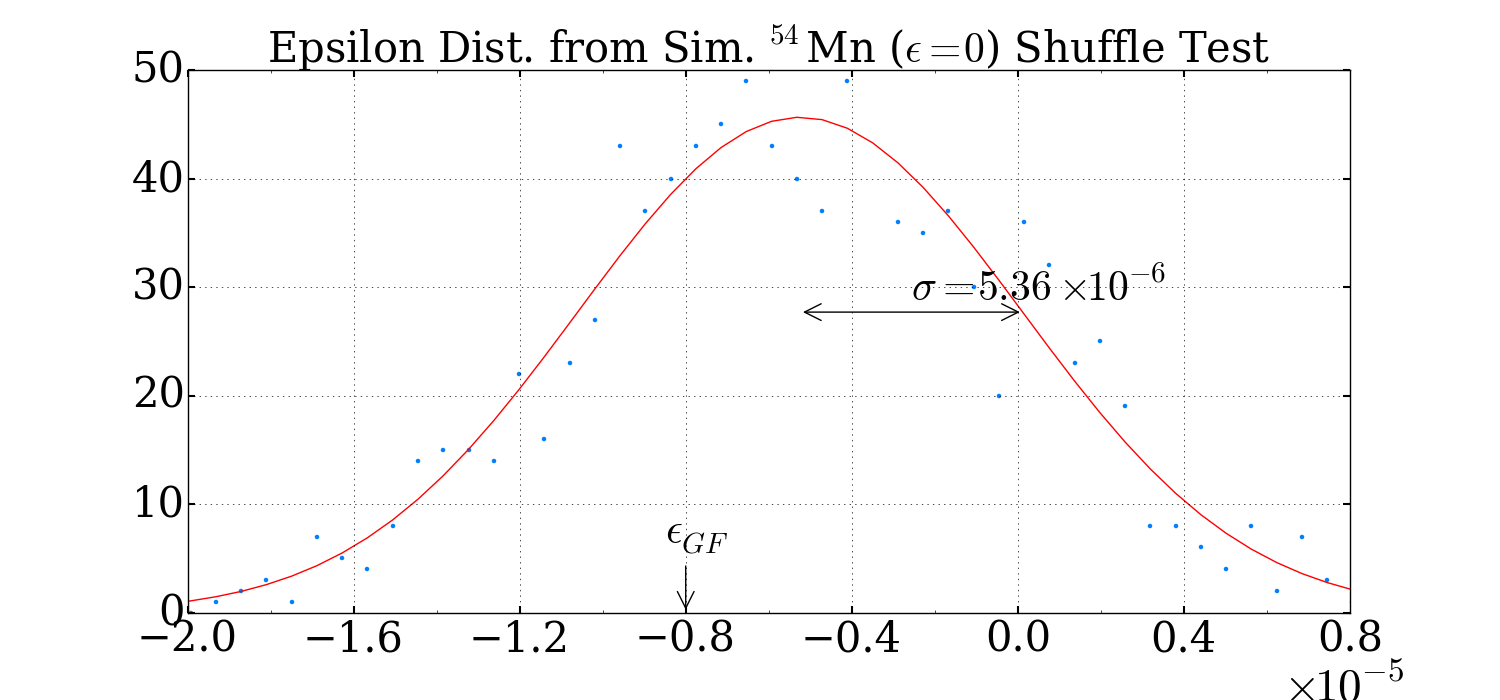}}
    
    \subfloat[]{\includegraphics[width=0.47\textwidth]{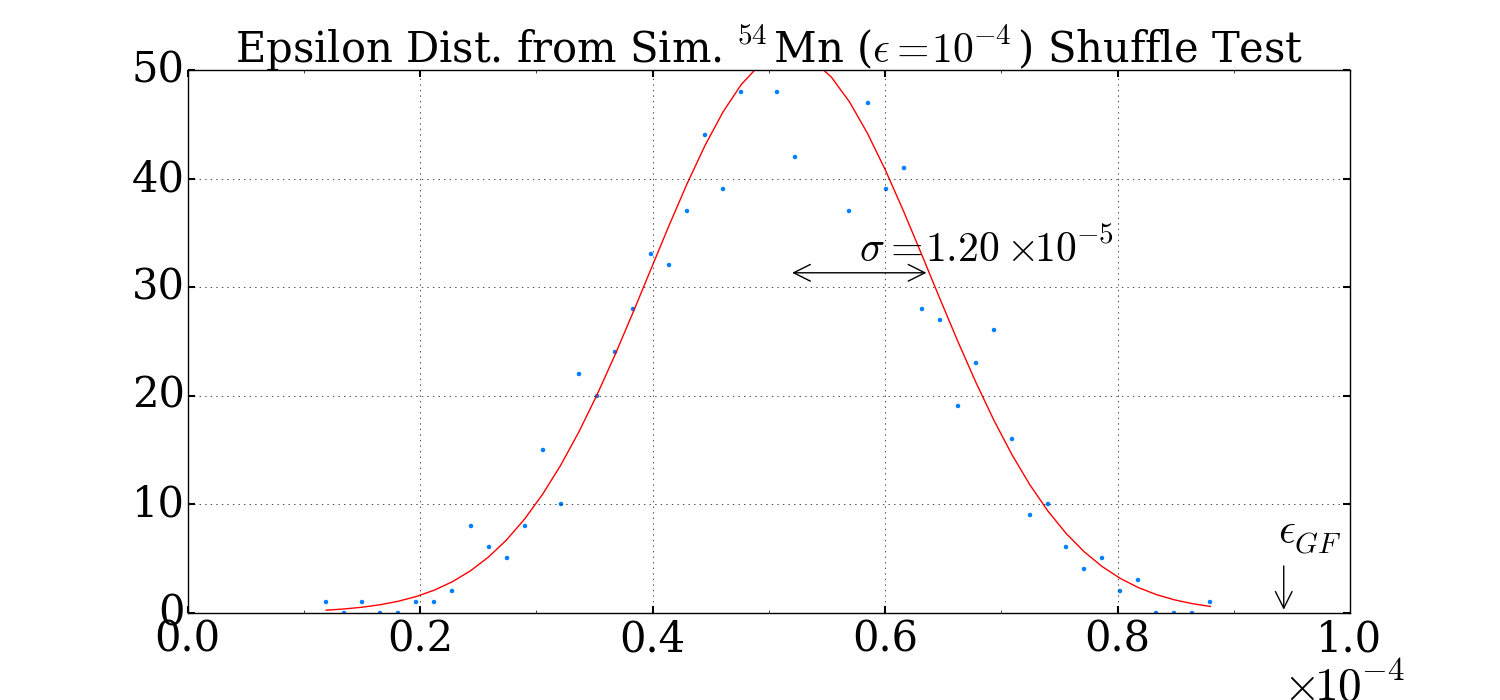}}
     \caption{Distributions of $\epsilon$ from Global Fits to 1000 Monte Carlo random shuffles of the reactor transition times for two simulated data sets: (a) $\epsilon$ = 0 (b) $\epsilon$ = $10^{-4}$.  Both use typical $^{54}$Mn counting rates [$3\times10^7$ hourly counts].}	\label{fig:SimShuffle}
    
\end{figure}

\begin{figure*}
   \centering
	\subfloat[]{\includegraphics[width=0.47\textwidth]{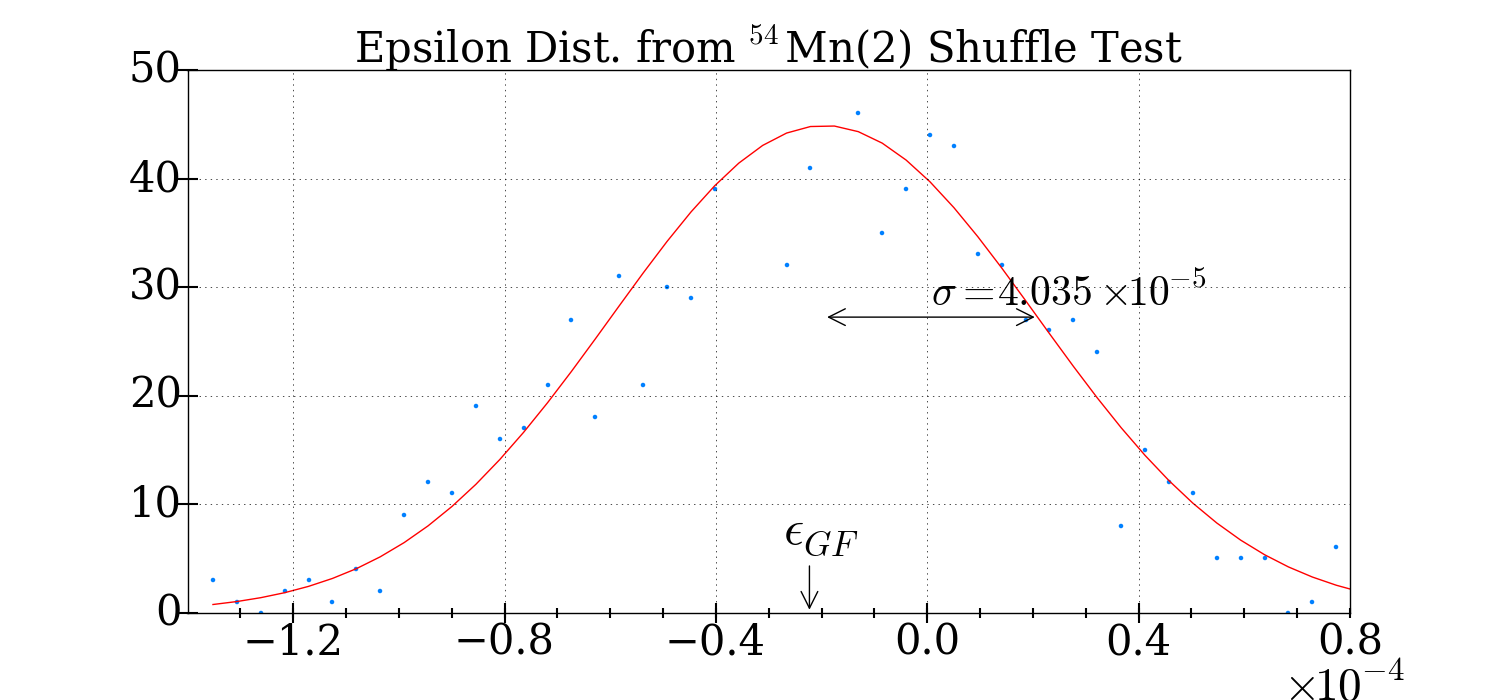}}
    \subfloat[]{\includegraphics[width=0.47\textwidth]{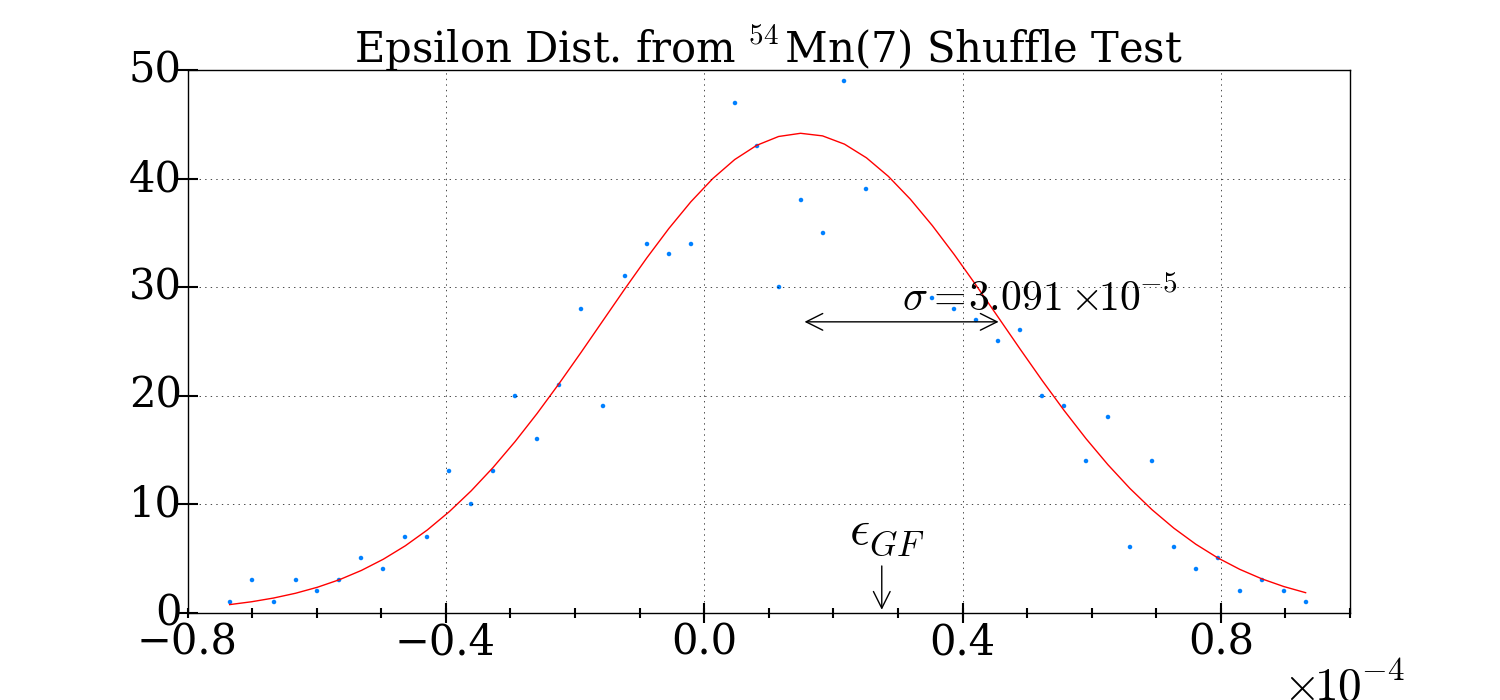}}
    
    \subfloat[]{\includegraphics[width=0.47\textwidth]{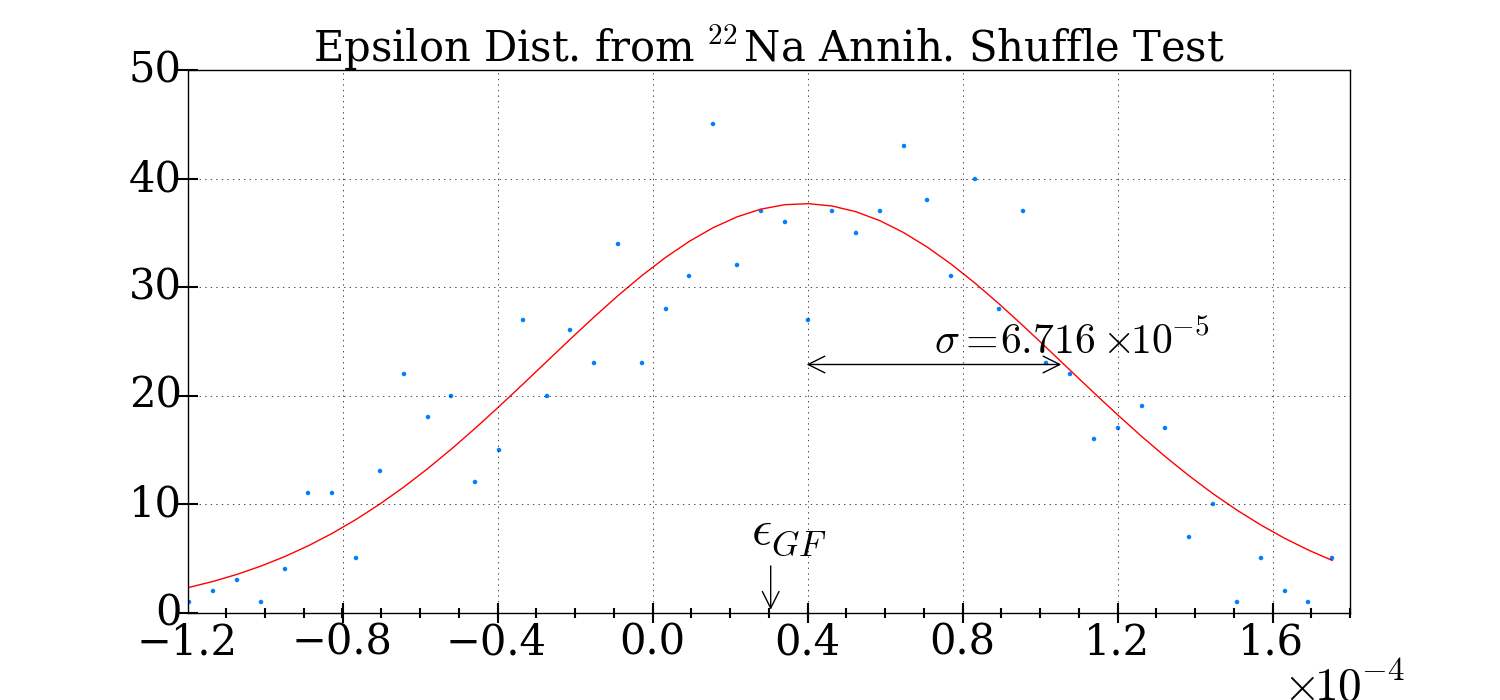}}
    \subfloat[]{\includegraphics[width=0.47\textwidth]{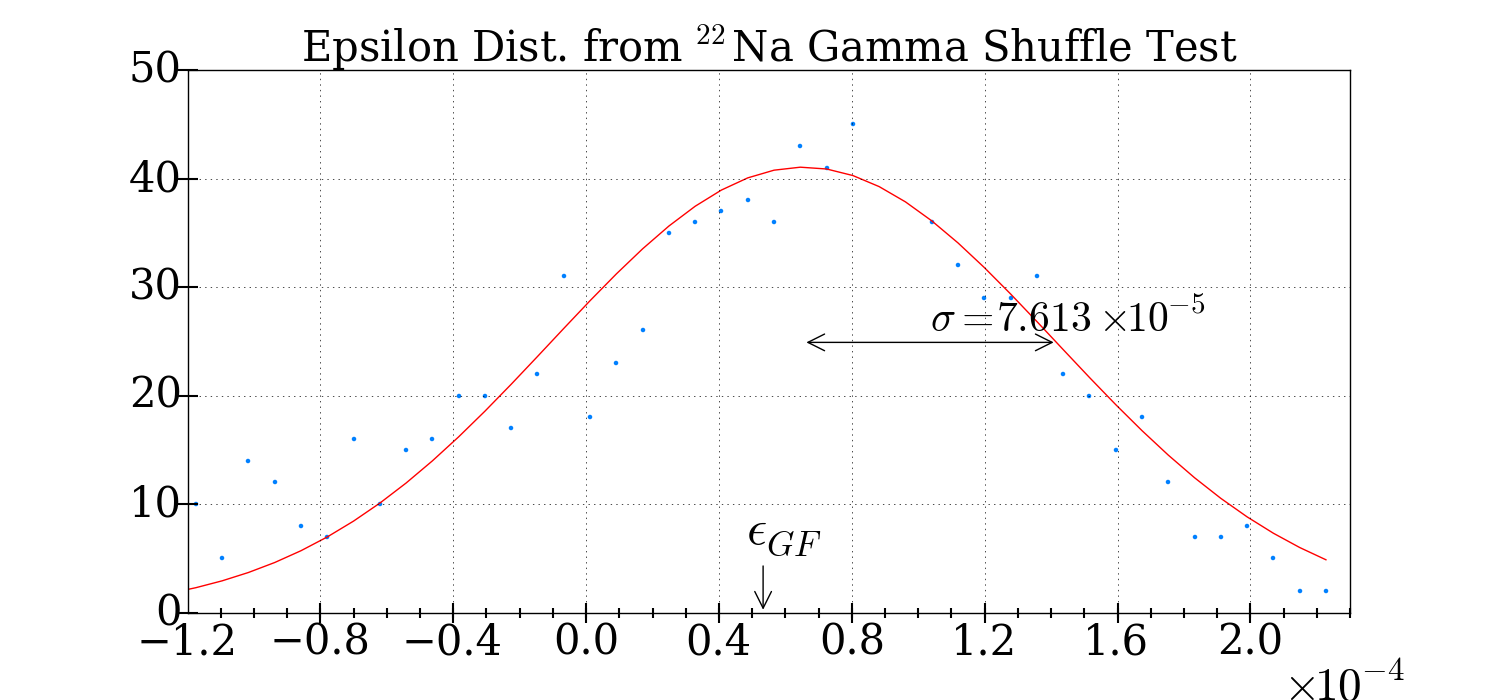}}
    
    \subfloat[]{\includegraphics[width=0.47\textwidth]{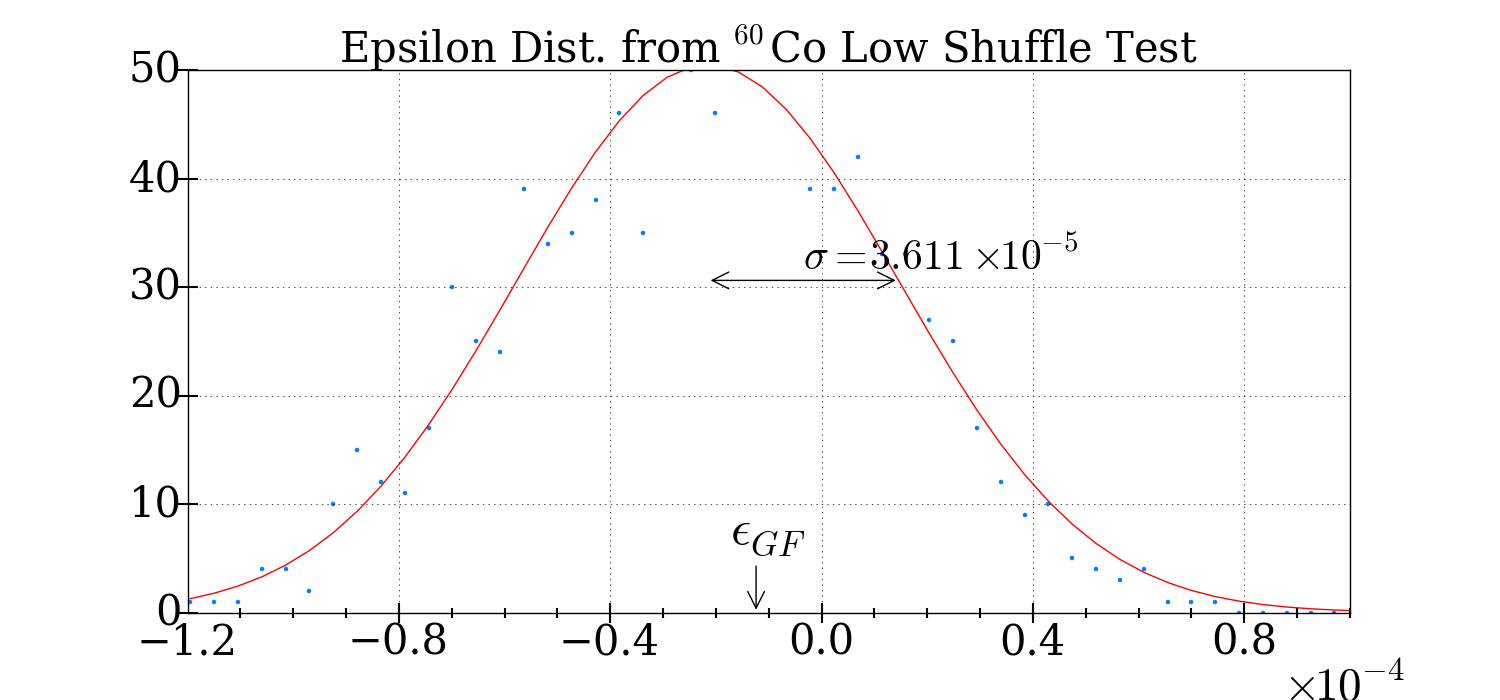}}
    \subfloat[]{\includegraphics[width=0.47\textwidth]{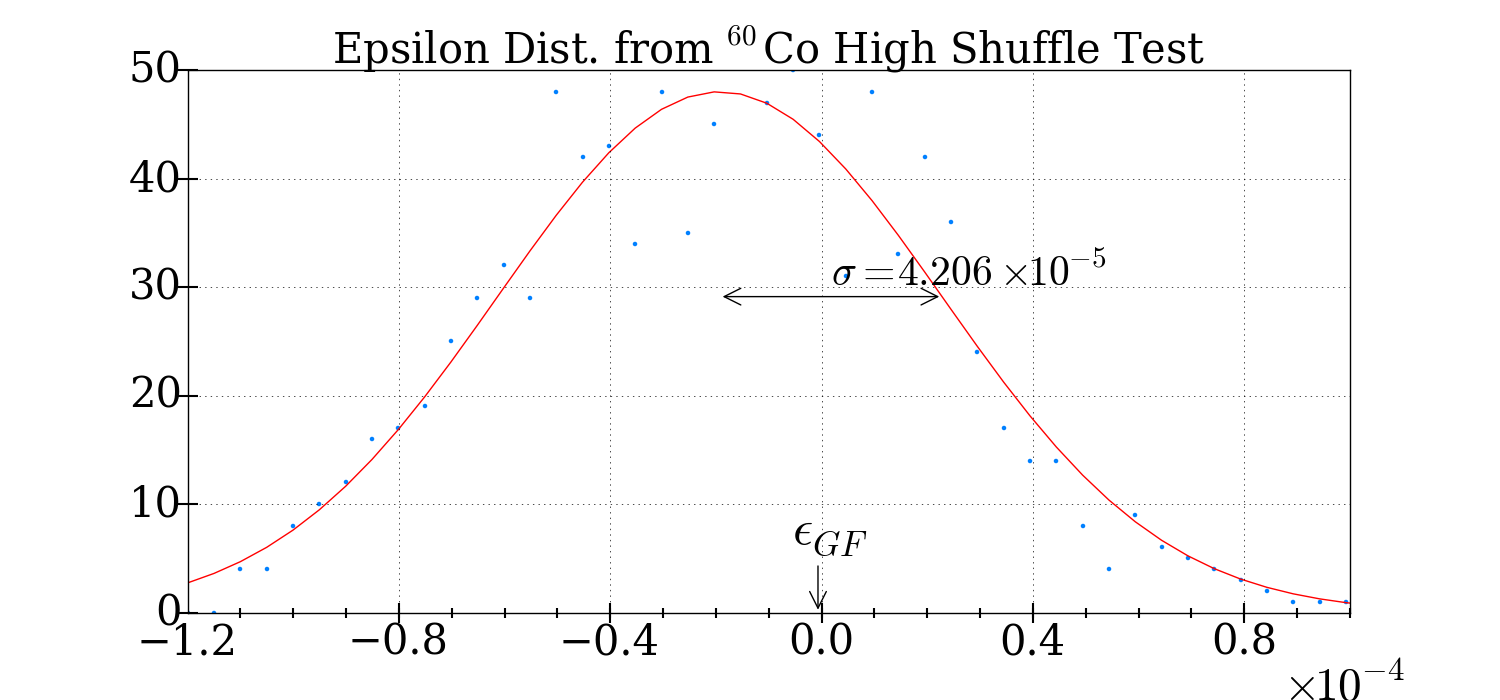}}
    \caption{Distributions of $\epsilon$ from Global Fits to 1000 Monte Carlo random shuffles of the reactor transition times for: (a) $^{54}$Mn Det.2 (b) $^{54}$Mn Det.7 (c) $^{22}$ Na annih. (d) $^{22}$Na gamma (e) $^{60}$Co low (f) $^{60}$Co high.}	\label{fig:Shuffle}
    
\end{figure*}

The Shuffle Test randomly shifts the \textsc{on}/\textsc{off} transition times from the actual values.  Each transition time is shifted randomly with uniform weight in a restricted interval. The first interval is from the start of the entire data set to the midpoint between the first and second transition times; the second interval extends from this midpoint to the next one, etc.  For each random choice of transition times, a Global Fit is performed.  This is repeated 1000 times and the distribution of $\epsilon$ values is plotted.  Since in general the shifted transition times treat a fraction of the \textsc{on} state as \textsc{off}, and \textsc{off} as \textsc{on}, any genuine reactor effect on the decay rate would tend to be diluted, driving the value of $\epsilon$ closer to zero, or even to the opposite sign if a there is a preponderance of wrong assignments of both reactor states.  In general, one would \textit{not} expect the magnitude of $\epsilon$ from a genuine reactor effect to be increased by the shuffle; $\sqrt{N}$ Gaussian statistics alone cannot drive values of $\epsilon$ larger than the best fit value in a shuffle test.  To demonstrate this, two simulated decay sets were generated with a $\lambda$ equal to that of $^{54}$Mn, one with no steps and one with $\epsilon$ = 1 $\times10^{-5}$    The statistics were determined by Gaussianly distributing the counts about the exponential, with $\sqrt{N}$ statistics and $N_0$ equal to $3\times 10^{7}$ counts per hour. The resulting data streams were then processed by the Global Fitting procedure.  The dilution of $\epsilon$ by incorrect transition times is clearly seen in Fig. \ref{fig:SimShuffle}; for the non-zero-$\epsilon$ case,  $\epsilon$ is never more than the input value (the no-step sequence has a "shuffle sigma" much smaller than all the others, reflecting the absence of systematic errors).  Figure \ref{fig:Shuffle} shows the frequency distributions of $\epsilon$ for the three isotopes. All six distributions are reasonably fitted by Gaussians.  For the two $^{60}$Co peaks, the values of $\epsilon$ are widely distributed on either side of the (small) Globally Fitted values, with standard deviations ($\sigma$) of 3.6$\times10^{-5}$ and 4.2$\times10^{-5}$, which we take to be estimates of the effects of the visible systematic fluctuations in the de-trended data. Note that these standard deviations are considerably larger than the Globally Fitted statistical uncertainties.  The $^{22}$Na shuffle distributions, in more than half of the cases, \textit{increase} the values of $\epsilon$, clearly impossible for a genuine reactor effect.  The Gaussian $\sigma$'s are 6.9$\times10^{-5}$ and 7.9$\times10^{-5}$, again considerably larger than the Globally Fitted statistical errors. For the $^{54}$Mn in Detector 2, the value of $\sigma$ is 3.0$\times10^{-5}$ and many values of $\epsilon$ are greater than the Globally Fitted value, comparable to the other isotopes.  For Detector 7 the $\sigma$ from the Shuffle Test is 1.6$\times10^{-5}$ and in a significant number of cases $\epsilon$ is greater than the Globally Fitted value---again incompatible with a genuine reactor effect.\\

As discussed above, some of the systematic error estimates comes from the differences in two values of $\epsilon$: the "Fit Change"  from the two alternative Global Fits, for the three isotopes; and the large Mn $\Delta\epsilon$ discrepancy.  Using the "scatter" of two values, we apply the formula \cite{wiki:standarddev} for the corrected and unbiased estimator of $\sigma$, from two samples:  

\begin{equation} \label{eq:sigmafromtwosamples.1}
 \sigma = \frac{s}{C_4(N)}.
\end{equation} 
With
\begin{equation} \label{eq:sigmafromtwosamples.1}
s = \sqrt{\frac{\sum(x_i - \overline{x})^2}{N - 1}}\ \ \textrm{and} \ \ C_4(N) = \sqrt{\frac{2}{N-1}}\frac{\Gamma(\frac{N}{2})}{\Gamma(\frac{N-1}{2})},
\end{equation} 

\begin{equation} \label{eq:sigmafromtwosamples.1}
 \textrm{it follows that} \ \  \sigma = \frac{|x_1 - x_2|}{1.128} \ \ \textrm{for} \ \ N = 2.
\end{equation} 

This gives $\sigma$ = $|\epsilon_{Det7} - \epsilon_{Det2}|$/1.128 = $4.66\times10^{-5}$ for the two Mn detectors, for example.\\

\begin{table*}
\caption{Epsilons with All Errors}
\label{tab:allepsilons}
\setlength{\extrarowheight}{3pt}
\begin{ruledtabular}
\begin{tabular}{ccccccccc}

&&&&&\multicolumn{4}{c}{$\pm$ syst.}\\
\cline{6-9}
Isotope & Feature & 			$\epsilon$ & 		$\pm$ stat. 		& $\Delta \epsilon$ & Bkgd. 		& Fit Change 			& Shuffle 				& $\Delta \epsilon$ \\
\hline
$^{54}$Mn & Det. 2		& -2.23$\times10^{-5}$	&  6.1$\times10^{-6}$ 	& 	&2.6$\times10^{-6}$ & 9.1$\times10^{-6}$ &4.04$\times10^{-5}$ & \\
          & Det. 7		& 2.75$\times10^{-5}$	& 5.6$\times10^{-6}$	&					&2.3$\times10^{-6}$ & 1.2$\times10^{-5}$&3.09$\times10^{-5}$ &          \\
          & Wtd. Avg.	& 4.80$\times10^{-6}$	& 4.1$\times10^{-6}$ 	& 	5.98$\sigma$					&3.5$\times10^{-6}$ & 1.5$\times10^{-5}$ & 5.08$\times10^{-5}$&   4.42$\times10^{-5}$      \\
          \hline
$^{22}$Na & 511 keV 	&  3.05$\times10^{-5}$ 	& 6.1$\times10^{-6}$	& 	&4.2$\times10^{-6}$ & 7.9$\times10^{-7}$&6.72$\times10^{-5}$ &  \\ 
          & 1275 keV 	&  5.33$\times10^{-5}$ 	& 1.5$\times10^{-5}$ 	& 					&9.0$\times10^{-6}$& 8.6$\times10^{-6}$&7.61$\times10^{-5}$ &  \\
          & Wtd. Avg. 	&  3.38$\times10^{-5}$	& 5.6$\times10^{-6}$ 	& 	1.41$\sigma$					&1.0$\times10^{-5}$ & 4.7$\times10^{-6}$&7.43$\times10^{-5}$ & (2.02$\times10^{-5}$ )\\
          \hline
$^{60}$Co & 1173 keV 	&  -1.24$\times10^{-5}$	& 8.3$\times10^{-6}$ 	& 		&3.0$\times10^{-6}$ & 1.9$\times10^{-7}$ &3.61$\times10^{-5}$ &  \\
          & 1333 keV 	&  6.82$\times10^{-7}$ 	& 9.4$\times10^{-6}$ 	&					&3.3$\times10^{-6}$ & 7.6$\times10^{-7}$&4.21$\times10^{-5}$ &    \\
          & Wtd. Avg. 	&  -7.24$\times10^{-6}$ 	& 6.2$\times10^{-6}$ 	& 	0.93$\sigma$			&4.4$\times10^{-6}$ & 4.7$\times10^{-7}$&3.90$\times10^{-5}$ &   (1.04$\times10^{-5}$)  \\

\end{tabular}
\end{ruledtabular}
\end{table*}

The values of $\epsilon$ from the six peaks are given in Table \ref{tab:allepsilons}, together with their statistical errors and the four types of systematic error discussed above.  Also shown are the statistically weighted average of the two $\epsilon$ values for each isotope, and the statistical significance of the differences in each pair of $\epsilon$ values.   The background systematic errors are propagated in quadrature for each isotope.  The differences in $\epsilon$ for $^{22}$Na and $^{60}$Co are not statistically significant, which is not surprising since the same detector is measuring both peaks for the isotope.  Note  that the Shuffle Test systematic errors are quite similar within a given detector.  The Fit Change estimates for  $^{22}$Na and $^{60}$Co are small and will contribute little to the overall uncertainty.  The 1275 keV Na data also have small statistical weight relative to the more abundant positron annihilation peak .  Since some of the unknown systematic errors are probably common to a given detector, we use the mean of the two Shuffle uncertainties in the summary lines for $^{22}$Na and $^{60}$Co in Table \ref{tab:allepsilons}, and similarly for the Fit Change column. Further justification for this choice can be seen in parts (c),(d) of Fig. \ref{fig:AlphaDetrended}: the long-term features of the $^{22}$Na data for the two peaks are similar, showing some degree of correlation.  In parts (e),(f) of this figure, for the two $^{60}$Co peaks, the shorter term fluctuation patterns appear rather similar, again showing some degree of correlation. The case is not so clear for $^{54}$Mn, since two different detectors are involved; but the values of the uncertainties in each of the two columns are nearly equal. Moreover, the $^{54}$Mn parts of the figure,  (a),(b) show some degree of correlation, with an initial drop followed by a dip at around day 70 in both detectors (the dip at around day 70 is also present for the Co plots (e),(f)).  In any case,  we conservatively propagate the Fit Change and Shuffle errors for $^{54}$Mn in quadrature.\\

\section{Conclusions}

Table \ref{tab:finalepsilons} summarizes the three values of $\epsilon$ from the three isotopes.  The statistical errors, and the small systematic errors in relative bay-to-bay background levels, are both almost insignificant when added in quadrature to each of the estimates of unknown systematic errors for each isotope (note that because there is no significant discrepancy, the  $\Delta\epsilon$ estimates for $^{22}$Na and $^{60}$Co are not used). We quote the final results for $\epsilon$ as two-sided 95$\%$ C.L. "outer limits" (mean value $\pm$ 2$\times$ overall $\sigma$).\\ 

 In summary, this experiment, while quite sensitive, cannot exclude perturbations less than one or two parts in 10$^4$ in $\beta^{\pm}$ decay (or electron capture) processes, in the presence of an antineutrino flux of 3$\times$10$^{12}$ cm$^{-2}$s$^{-1}$. Our 10$^{-4}$ sensitivity level is roughly comparable to that of de Meijer and Steyn \cite{deMeijer2}, who find for $^{22}$Na a relative change between reactor-\textsc{on} and reactor-\textsc{off} of ($-$0.51$\pm$ 0.11)$\times10^{-4}$, but do not exclude possible systematic errors. If we assume that the interaction strength of $\bar{\nu_e}$ with radioactive isotopes is the same as that of $\nu_e$, then our upper limit excludes $\nu_e$ as the dominant source of the $\mathcal{O}(10^{-3})$ effects reported in some of the original papers \cite{1,2,3,4,5,6,7,8,9,10,11,12,13,14,15,16,17,18,19,20,21,22,23,24,25}. However, as noted in the Introduction, our results do not impose any constraints on the coupling strengths of other neutrino flavors, which are present in both the solar neutrino flux, and as components in the cosmic neutrino background, but not significantly elevated 6 m from the HFIR reactor core. The present experimental methods are applicable to a wide variety of radioactive isotopes. An improvement in sensitivity would be possible if we could understand and reduce the dominant systematic uncertainties, which are presently of unknown origin(s). 

\begin{table*}
\caption{Final Epsilons}
\label{tab:finalepsilons}
\setlength{\extrarowheight}{3pt}
\begin{ruledtabular}
\begin{tabular}{ccccc} 

 Isotope  & $ \epsilon $   & Total error & 95$\%$ LL & 95$\%$ UL \\
\hline
$^{54}$Mn &  4.80$\times10^{-6}$  &  6.91$\times10^{-5}$  &-1.33$\times10^{-4}$ & 1.43$\times10^{-4}$ \\

$^{22}$Na &  3.38$\times10^{-5}$  &  7.27$\times10^{-5}$  & -1.12$\times10^{-4}$ & 1.79$\times10^{-4}$  \\

$^{60}$Co &  -7.24$\times10^{-6}$  & 3.98$\times10^{-5}$ & -8.69$\times10^{-5}$ & 7.24$\times10^{-5}$ \\

\end{tabular}
\end{ruledtabular}
\end{table*}



\appendix
\section{The Global Fitting Formalism}
\label{app:GlobalFitting}

The Global Fitting formulas described in Section VI uses just four parameters:  $C_0$ (the initial counts in one hour); the decay constant, $\lambda_0 \equiv \lambda_{\textsc{off}}$; $\epsilon \equiv (\lambda_{\textsc{on}}-\lambda_{\textsc{off}})/\lambda_{\textsc{off}}$; and the parameter $\alpha$ used in the rate-dependent distortion factor described in Section V.  Clearly $\lambda_{\textsc{on}} = \lambda_{\textsc{off}}(1+\epsilon)$.  We measure the counting rates in the form of the dimensionless number of counts in each one hour of live time.  For clarity in relating the number of remaining radioactive nuclei to the counting rate, we assume 100$\%$ counting efficiency of the decaying nuclei.\\

To illustrate the functional form of the Global Fit, assume that an experiment began in an \textsc{off} period and lasted for a time $T_{1}$, followed by an \textsc{on} period which lasted for a time $T'_{1}$. Then at the end of the first \textsc{on} and \textsc{off} periods we have for $\textsc{off}_{1}$\\

\begin{equation} \label{eq:A.2}
N(T_1)=N_0e^{-\lambda_0T_1}
\end{equation} 
and for $\textsc{on}_1$,
\begin{equation} \label{eq:A.3}
N(t)=N(T_1)e^{-\lambda_0(1+\epsilon)(t-T_1)}
=N_0e^{-\lambda_0[t+\epsilon(t-T_1)]}
\end{equation} 

\begin{equation} \label{eq:A.5}
N(T_1+T'_1)=N_0e^{-\lambda_0[(T_1+T'_1)+\epsilon T'_1]}.
\end{equation} 

The first term of the exponential in Eq.(\ref{eq:A.5}) gives the surviving number of atoms after a cumulative time $(T_1+T'_1)$ starting from $N_{0}$, assuming no influence from reactor antineutrinos. For $\epsilon>0$ assumed, the $\epsilon-$dependent contribution then gives an additional loss during the time $T'_1$ due to the excess in the decay constant, $\lambda_0\epsilon$. The results in Eqs.(\ref{eq:A.2}-\ref{eq:A.5}) can be generalized in an obvious way: the number of surviving atoms $N(t)$ after a cumulative elapsed time $t$ is given by\\

\begin{equation} \label{eq:A.6}
N(t)=N_0e^{-\lambda_0(t+\epsilon\sum_{i} {T'_i})}
\end{equation}

In Eq.(\ref{eq:A.6}), $\sum_{i} {T'_i}$ extends over only \textsc{on} periods $T'_i$, during which it is assumed that any additional reactor-\textsc{on} contributions enter with the same factor $\epsilon$. For the \textsc{hfir} reactor at Oak Ridge this assumption is justified given that during each \textsc{on} period the reactor runs at the same 85 MW rate using the same fuel composition. \\

What is actually measured is the counting rate, or its proxy, the hourly count total $C(t)$.  Incorporating the counting rate dependent distortion factor described in Section VI, $e^{\alpha C(t)}$ gives\\

for \textsc{off} periods:\\

\begin{equation} \label{eq:A.7a}
C(t)=C_0e^{-\lambda_0(t+\epsilon\sum_{i} {T'_i})+\alpha  C(t)}
\end{equation}

and for \textsc{on} periods:\\

\begin{equation} \label{eq:A.7b}
C(t) = C_0(1+\epsilon)e^{-\lambda_0(t+\epsilon\sum_{i} {T'_i})+\alpha  C(t)}.
\end{equation}

Once again, $\sum{T'_i}$ is the sum of all \textsc{on} time intervals up to the data point at time $t$.


\section{Background Subtractions}
\label{app:BackgroundSubtraction}
As can be seen in Figs \ref{fig:Detrended} and \ref{fig:AlphaDetrended}, the backgrounds are low and steady in the reactor-\textsc{off} periods, but in the reactor-\textsc{on} periods are larger and highly irregular due to several intermittently-operated neutron beam lines one floor below the detectors.  The general effect of such backgrounds, if not dealt with, would mimic an increased decay rate during reactor-\textsc{on} periods, and could generate false steps and a spurious positive value of $\epsilon$.  To deal with these backgrounds, two counters, with no installed sources, were located in bays $\#1$ and $\#8$, on the lower and upper levels, respectively, of the cave---as in Table \ref{tab:detector-layout}. We use the background information on an hour-by-hour basis.
 
At the start of Phase II, which was during a reactor-\textsc{on} period, each of the two background counters was placed in each of the eight pockets (bays) of the lead "cave", to measure the relative background levels. The results as measured by Detector 8 are shown in Table \ref{tab:scalepercentages}, where the background-level scale factors of the top row are relative to bay $\#8$  and the bottom row scale factors are relative to bay $\#1$. These original calibration runs had five minute live times, and in Detector 1 typically had about 4800 counts in top-row bays and 3500 counts in bottom-row bays. The resulting statistical uncertainties in Table \ref{tab:scalepercentages} are 2.0$\%$ for the top row and 2.4$\%$ for the bottom row. These uncertainties lead to one type of systematic error on $\epsilon$, shown in Table \ref{tab:allepsilons}, which is quantified by varying the amount of background subtracted for the Global Fit analysis.  The relative gains of all eight detectors are well measured and any gain-induced uncertainties will contribute negligibly to the sizes of the background subtractions.\\

\begin{table*}
\centering
\caption{Ratio of Initial Background Counts Relative to Detector 1.  See text for further discussion.}
\label{tab:scalepercentages}
\setlength{\extrarowheight}{3pt}
\begin{tabular}{|l|l|l|l|l|}
\hline
Det5 Am: 0.917   & Det6 Na: 0.871\footnotemark[2] & Det7 Mn: 0.951 & Det1 Bkg: 1.0000\footnotemark[1] & $\pm$ 0.020 \\
\hline
Det1 Bkg: 1.0000\footnotemark[1] & Det2 Mn: 0.895 & Det3 Co: 0.936 & Det4 Eu: 1.075 &  $\pm$ 0.024 \\
\hline
\end{tabular}
\footnotetext[1]{Top row ratios are relative to Pocket 8 and bottom row ratios are relative to Pocket 1}
\footnotetext[2]{For the Na annih. peak, background subtraction is done with Det. 8 data, and the initial ratio relative to Det. 8 is 0.911.}
\end{table*}

\begin{table*}
\centering
\caption{Gains relative to Detector 1 (2.646 keV/bin).  See text for further discussion.}
\label{tab:inversescalefactors}
\setlength{\extrarowheight}{3pt}
\begin{tabular}{|l|l|l|l|}
\hline
Det5 Am: 9.093    & Det6 Na:  1.326 & Det7 Mn: 2.412 & Det8 Bkg: 4.260 \\
\hline
Det1 Bkg: 1.000 & Det2 Mn: 1.478 & Det3 Co: 0.979 & Det4 Eu: 1.042 \\
\hline
\end{tabular}
\end{table*}

\begin{figure}
	\subfloat[]{\includegraphics[width=0.43\textwidth]{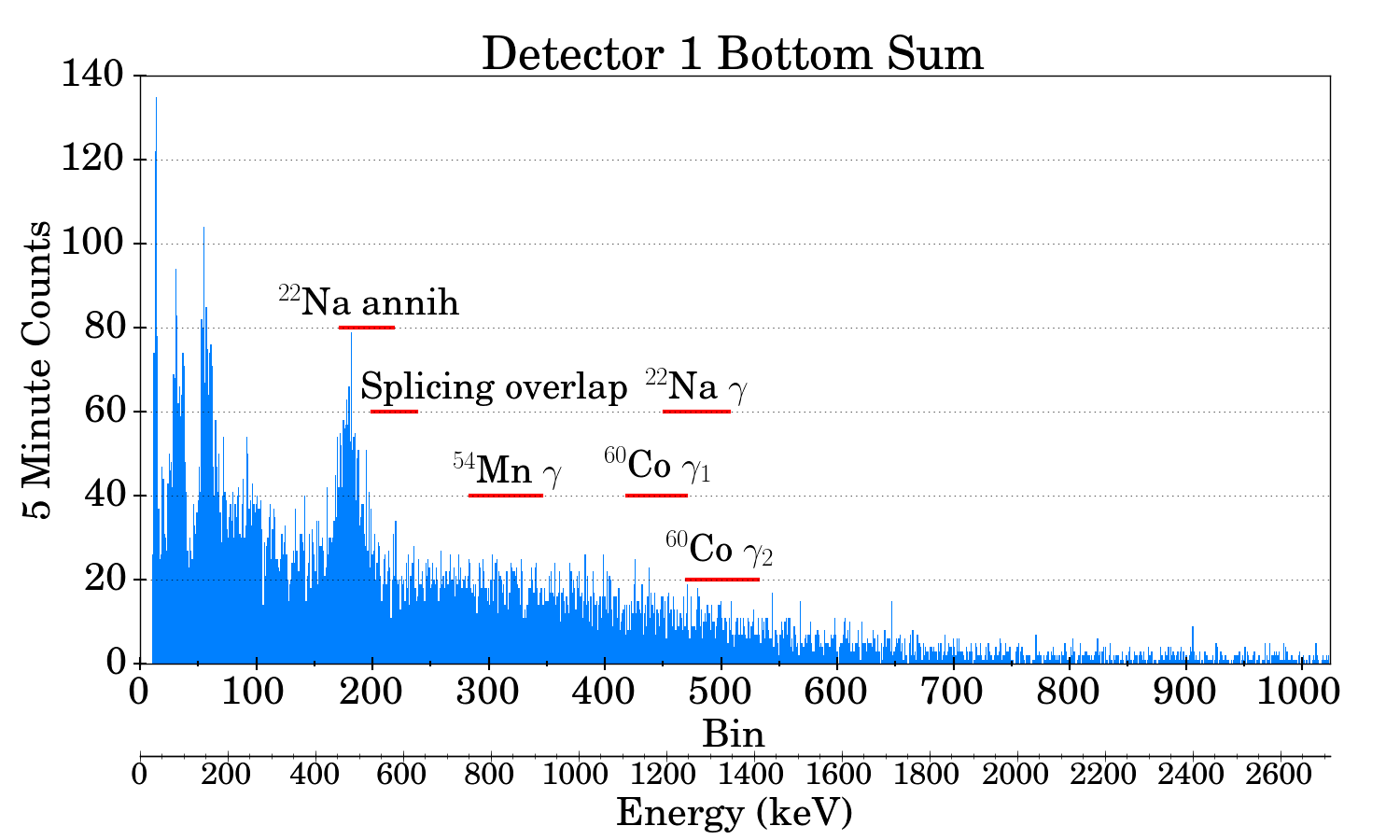}}
    
    \subfloat[]{\includegraphics[width=0.43\textwidth]{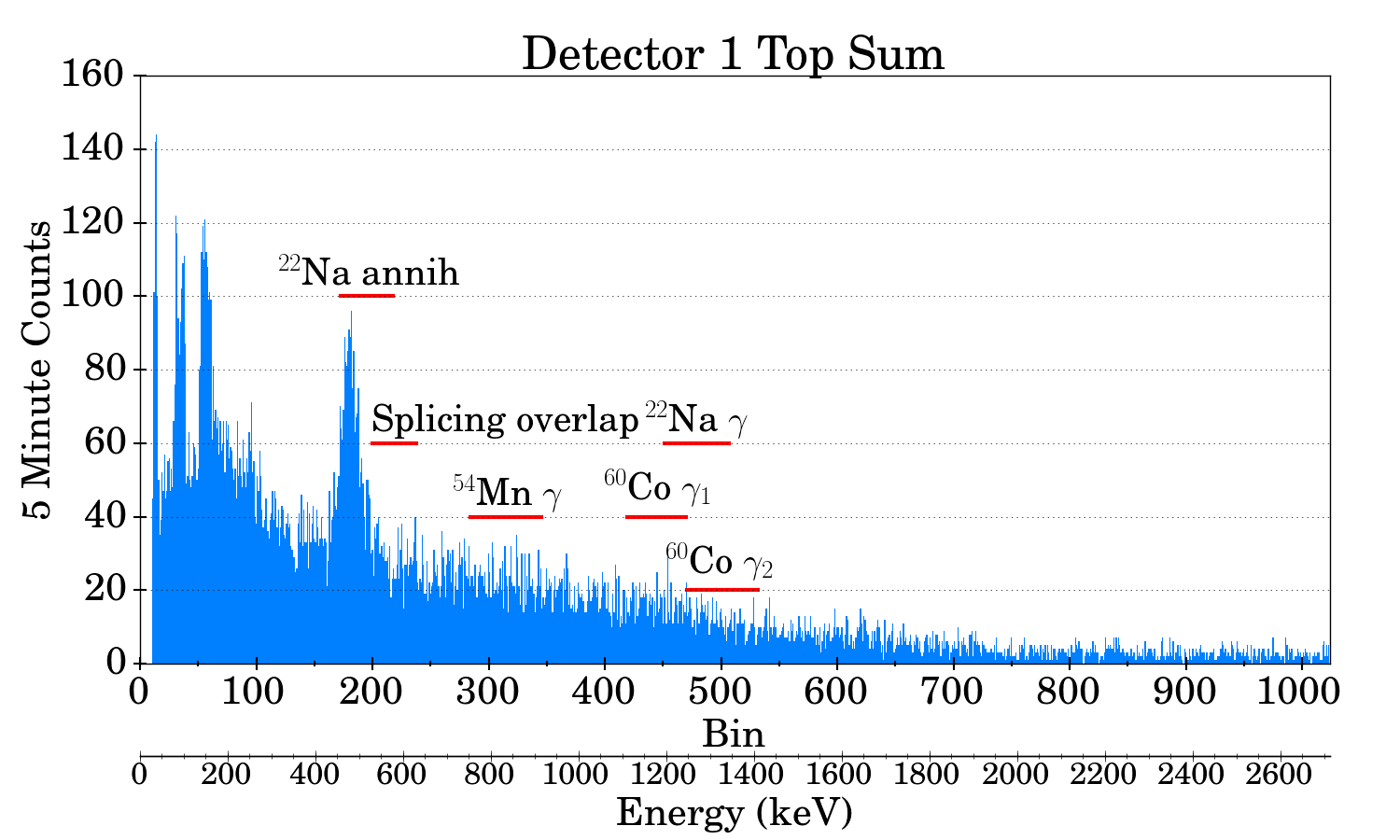}}
    
    \subfloat[]{\includegraphics[width=0.43\textwidth]{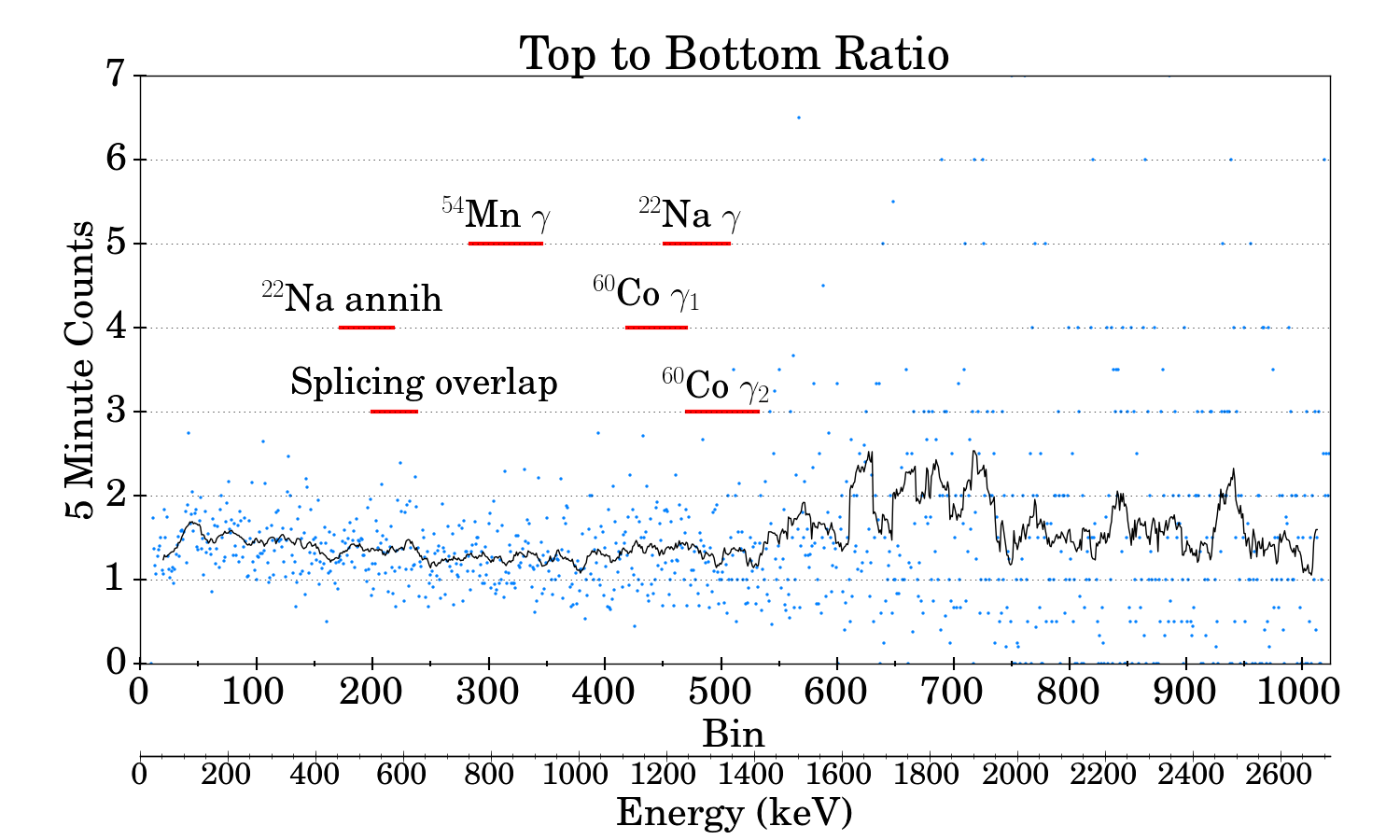}}
     \caption{Various sums of reactor-\textsc{on} spectra from 5 minute initial Phase II background runs with Detector 1 in all eight bays of the the cave: (a) Sum of top bays (b) Sum of bottom bays (c) Ratio of top sum to bottom sum.  Energy regions of interest and the Det1-Det8 scaling overlap region are shown as horizontal bars. The heavy black line is a 20 pt. moving average.}	\label{fig:SumBotTop}
    
\end{figure}

The lower four counters are better shielded than the upper-level  counters, by two lead layers instead of one. Also, central bays are more shielded than side bays. In general we expect the central bays to have somewhat smaller background counting rates than the side bays, and the top row of bays generally to have significantly larger background levels than the bottom row.  The gain of Detector 1 is set near or lower than any of the other gains, to cover all background energies that could lie at one of the source detectors' main gamma peaks.  The gain of Detector 8 is larger than any of the other gains (4.26 times the gain of Detector 1), and hence does not cover the energy range of almost all of the peaks of interest in counters with radioactive sources (with the exception of the $^{22}$Na positron annihilation peak).  Detector 8 can be used directly for the $^{22}$Na annihilation peak.  For the $^{22}$Na gamma peak and the Detector 7 $^{54}$Mn peak, We can approximately recover the missing Detector 8 information by using the Detector 1 information, scaled up by the ratio of counts in an overlap region defined by bins 860 to 1023 in Detector 8, which maps approximately onto bins 202 through 240 of the Detector 1 spectrum. Since the gains of the detectors are all different, we must of course map a given \textsc{roi} onto the corresponding energy range in the relevant background counter. And one must also take careful account of fractional bin contents. We now justify the the use of Detector 1 data to replace missing Detector 8 data:
The Detector 1 spectra measured in bays 1 through 4 (Bottom) are summed, and similarly for bays 5 through 8 (Top), and are shown in Fig. \ref{fig:SumBotTop}. The spectral shapes are very similar, both in the scaling region and also in all energy regions of interest, as is seen in the Top/Bottom ratio plot of this figure. \\ 

\begin{figure}
	\subfloat[]{\includegraphics[width=0.42\textwidth]{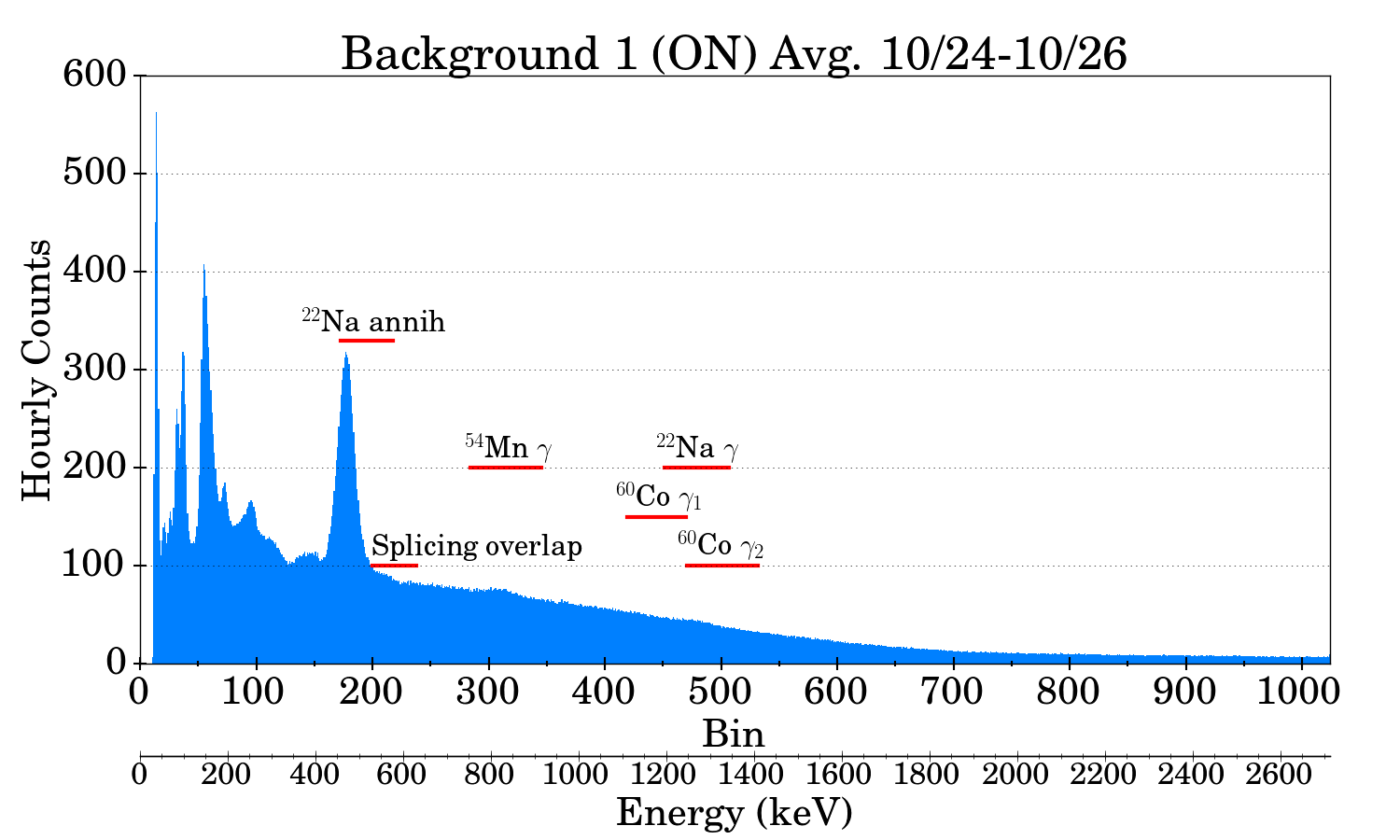}}
    
    \subfloat[]{\includegraphics[width=0.42\textwidth]{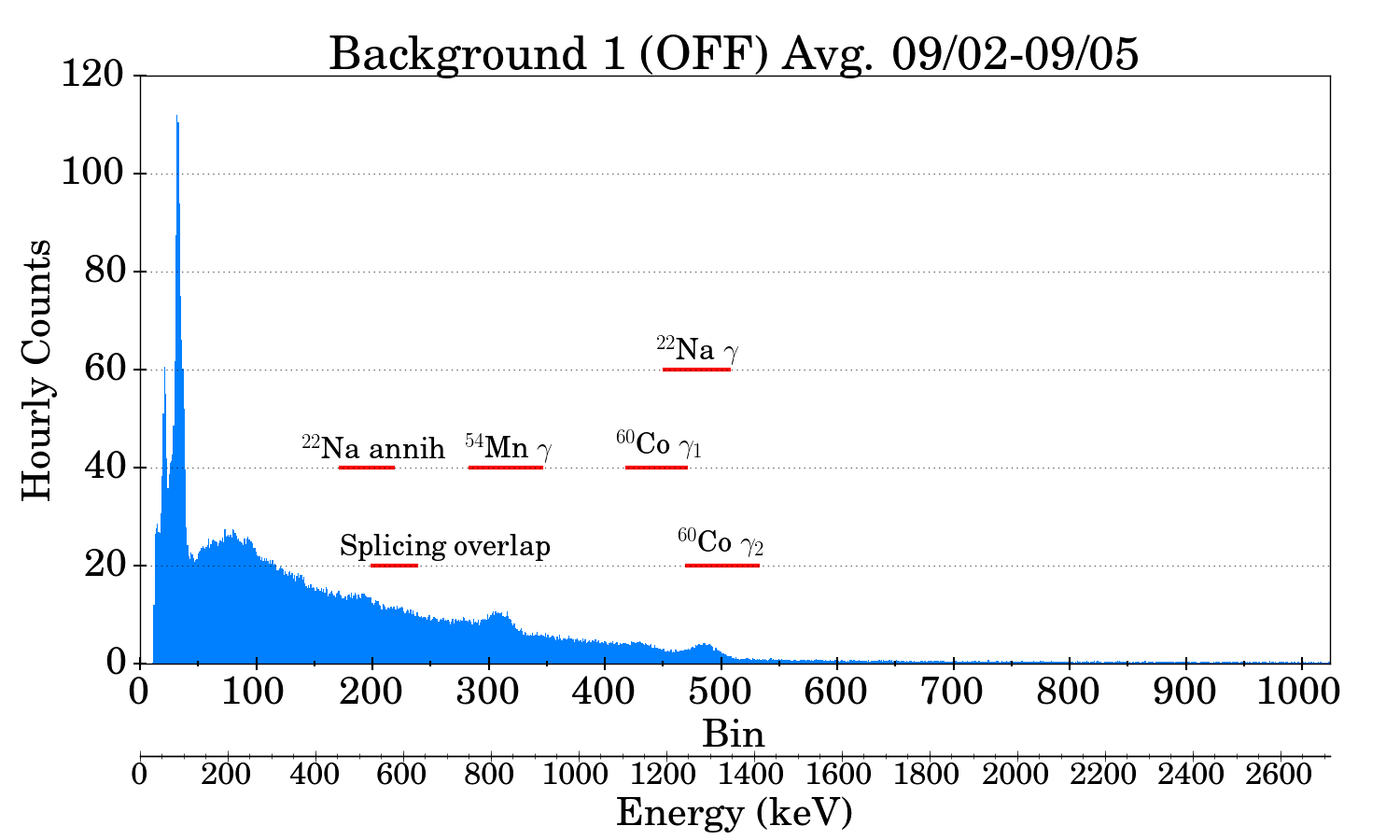}}
    
    \subfloat[]{\includegraphics[width=0.42\textwidth]{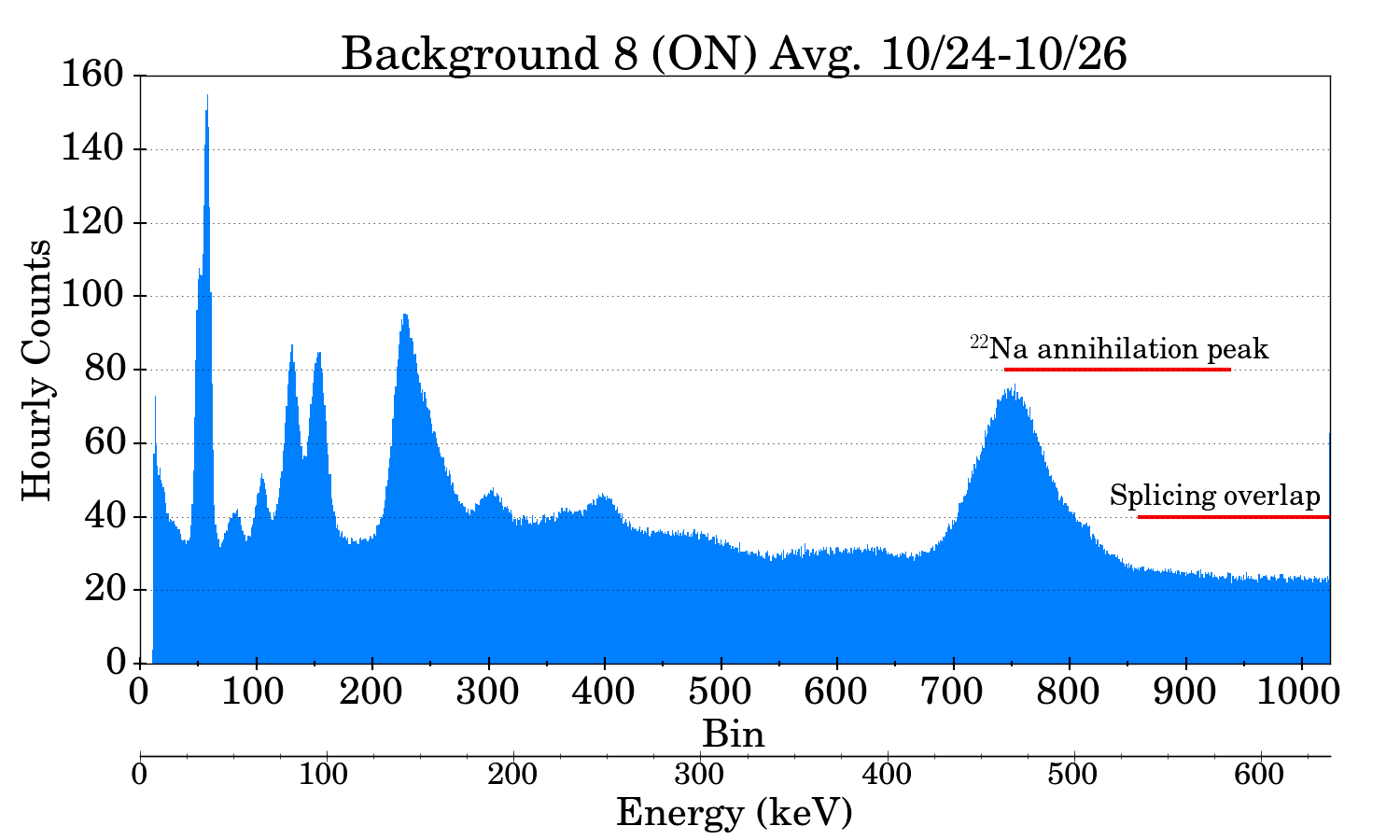}}
    
    \subfloat[]{\includegraphics[width=0.42\textwidth]{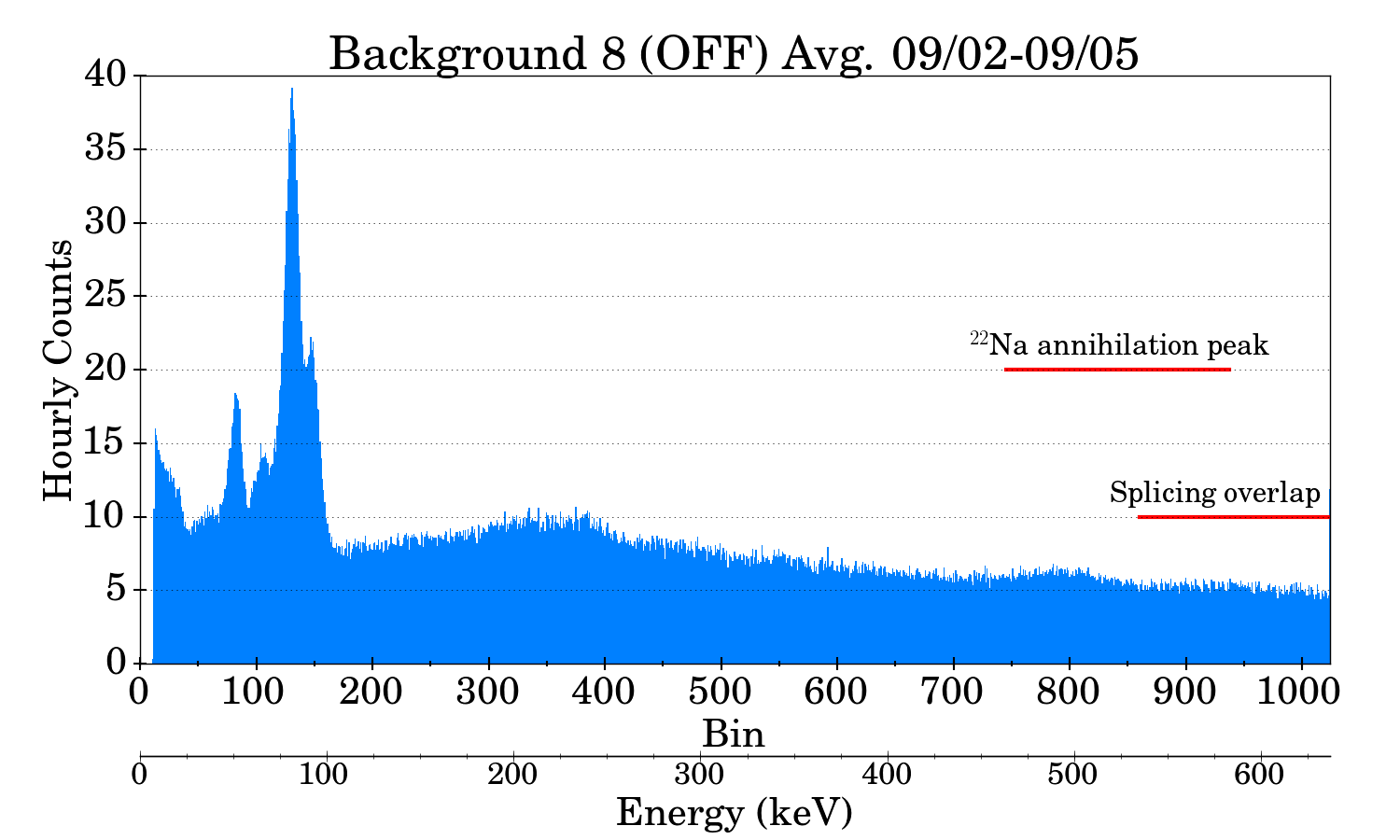}}
     \caption{3-day sums of background spectra for reactor-\textsc{on} and -\textsc{off} with Detector 1 in bay 1 and Detector 8 in bay 8: (a) Det. 1 \textsc{on} (b) Det. 1 \textsc{off} (c) Det. 8 \textsc{on} (d) Det. 8 \textsc{off}. Energy regions of interest and the Det1-Det8 scaling overlap region are shown as horizontal bars.}	\label{fig:3DaySumON}   
\end{figure}

From the main running periods, we have much higher background statistics by summing spectra from several days of running.  Also, this allows study of the reactor-\textsc{off} backgrounds. Such summed background spectra for both detectors for \textsc{on} and \textsc{off} are shown in Fig. \ref{fig:3DaySumON}. During \textsc{on} periods, the spectrum is rather smoothly decreasing after the last peak (which is somewhat below bin 200 in Detector 1). The chosen overlap scaling region is shown as a horizontal red bar, as are the various energy regions of interest. The \textsc{off} period background levels are some five times lower than for the ON period, and the spectrum falls off more quickly and has very little of the structure just below bin 200 in Detector 1. Hence the scaling/splicing procedure should work equally well, or better during \textsc{off} periods.\\ 

\begin{figure}
	\includegraphics[width=0.47\textwidth]{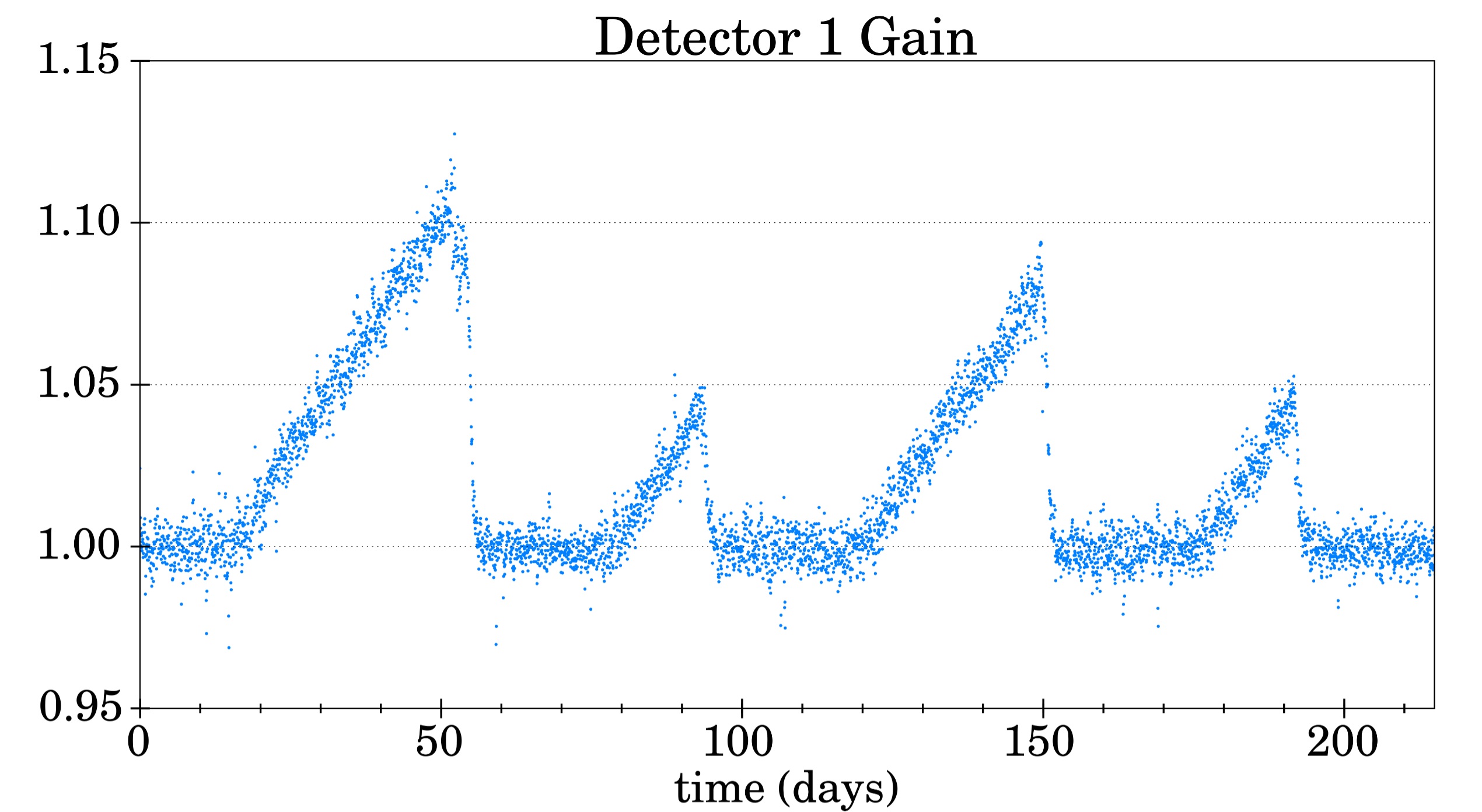}
     \caption{Hourly gain values of background Detector 1 vs. time, which are used to correctly map \textsc{roi}s onto the background spectrum.  Detector 8, not shown, had no gain variations.}	\label{fig:Det1Gain}
    \centering
\end{figure}

The Detector 1 gain locking was set for a peak in bin 54, which is present when the reactor is \textsc{on}.  This peak is absent when the reactor is \textsc{off}. The gain locking algorithm recognizes a peak in bin 13, and it very slowly and steadily (due to the very low statistics) increases the gain.  This can be seen in Fig. \ref{fig:Det1Gain}, where the gain increases by some 12 $\%$ over the first 32-day \textsc{off} period.  When the reactor turns \textsc{on}, the gain is restored fairly quickly (due to the five time higher counting rate).  This gain drift is easily corrected for in the subsequent analysis.  The Detector 8 gain locking worked perfectly, and no correction was needed.\\

The counters run asynchronously, with different deadtimes (and negligible deadtime for the background counters.)  Thus more than a single one-hour background count period is matched to a given source count period.  Background information is taken from these overlapping periods in such a way that only 60 minutes of \textit{live} background counts are considered.\\

\section{Energy Spectra and Decay Chains of Studied Isotopes}
\label{app:sources}

\begin{figure}
	\subfloat[]{\includegraphics[width=0.47\textwidth]{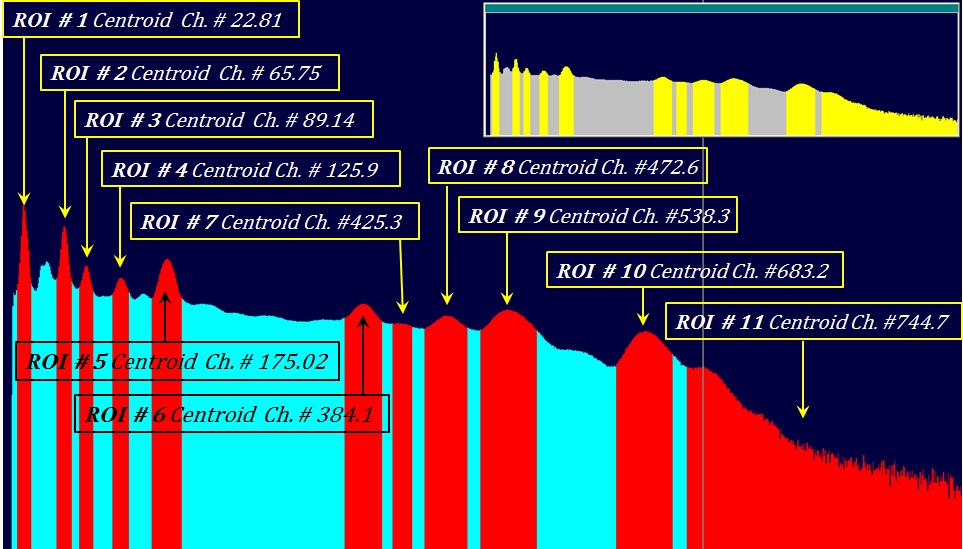}}
    
    \subfloat[]{\includegraphics[width=0.47\textwidth]{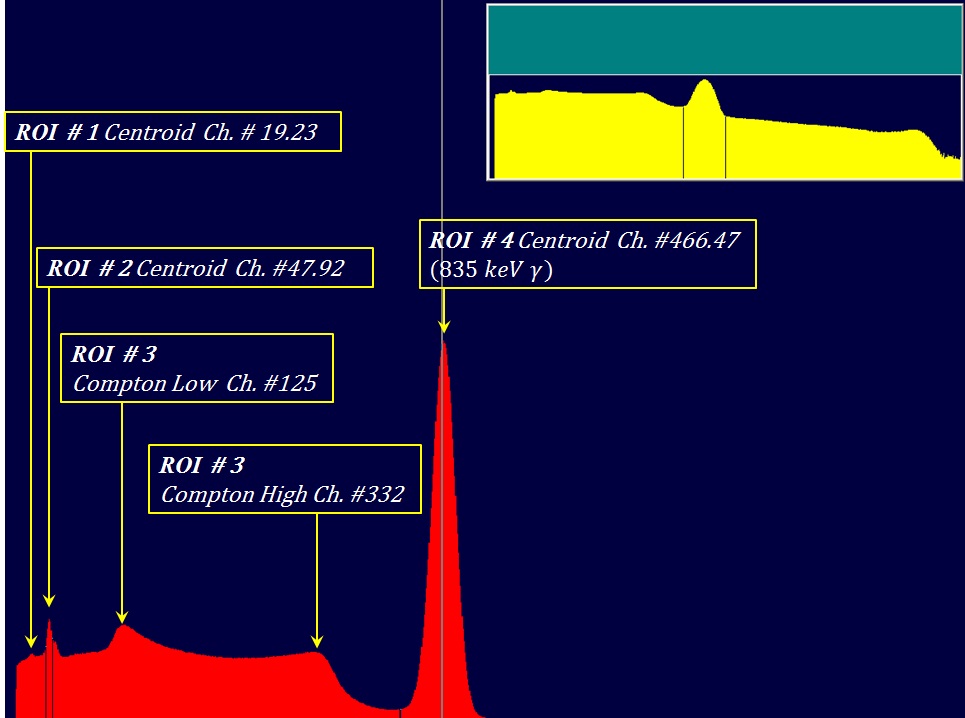}}
    
    \subfloat[]{\includegraphics[width=0.47\textwidth]{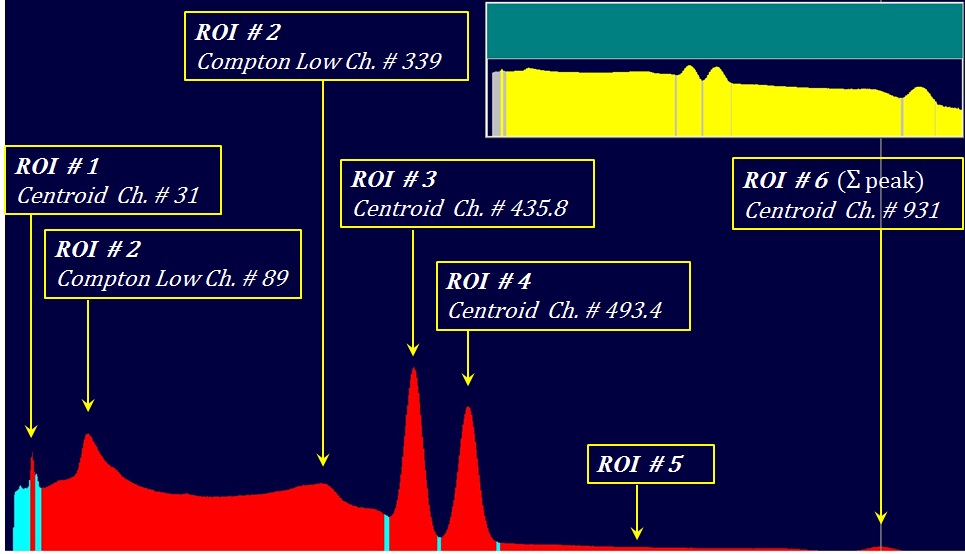}}

     \caption{Energy spectra with chosen \textsc{roi}s for: (a) $^{152}$Eu (b) $^{54}$Mn (c) $^{60}$Co.}	\label{fig:energyspectra}   
\end{figure}

Figure \ref{fig:energyspectra} contains \textsc{maestro} plots of the spectra from $^{152}$Eu, $^{54}$Mn, $^{22}$Na, and $^{60}$Co.  \textsc{roi}s are defined (in red) around each peak.  The decay chains of the isotopes are given in Table \ref{tab:decaychains}.

\begin{table*}
\caption{Decay Data of Sources Used in This Experiment}
\label{tab:decaychains}
\begin{flushleft}

$^{54}_{25}$Mn + $e^-$(EC) $\rightarrow$ $^{54}$Cr$^*$ + $\nu_e$ (100$\%$) ; $T_{1/2}$ = 312 d\\
\smallskip
\hspace{3.2cm}$\hookrightarrow^{54}_{24}$Cr(g.s.) + $\gamma$(834.8 keV) \\

\bigskip

$^{22}_{11}$Na $\rightarrow$ $^{22}_{10}$Ne + $e^{+}$ + $\nu_e$ (90$\%$) ; $T_{1/2}$ = 2.60 a = 949.7 d\\
\smallskip
\hspace{2.4cm}$\hookrightarrow$ $e^{+}e^{-}$ $\rightarrow$ $\gamma\gamma$(511 keV)\\

\bigskip

$^{22}_{11}$Na + $e^{-}$(EC) $\rightarrow$ $^{22}_{10}$Ne$^*$ + $\nu_e$ (10$\%$)\\
\smallskip
\hspace{3.2cm}$\hookrightarrow$ $^{22}_{10}$Ne(g.s.) + $\gamma$(1274.5 keV)\\

\bigskip

$^{60}_{27}$Co $\rightarrow$ ${}^{60}_{28}$Ni$^*$ + e$^-$ + $\overline{\nu}_e$ ($\sim$100$\%$) ; $T_{1/2}$ = 5.27 a = 1924.9 d\\
\smallskip
\hspace{1.6cm}$\hookrightarrow^{60}_{28}$Ni(g.s.) + $\gamma$(1173.3 keV) + $\gamma$(1332.5 keV)\\

\bigskip

$^{152}_{\phantom{0}63}$Eu + $e^{-}$(EC) $\rightarrow$ $^{152}_{\phantom{0}62}$Sm$^*$ + $\nu_e$ (73$\%$) ; $T_{1/2}$ = 13.54 a = 4945 d\\
\smallskip
\hspace{3.5cm}$\hookrightarrow$ $^{152}_{\phantom{0}62}$Sm(g.s.) + $\gamma$(122 keV)\\

\bigskip

$^{152}_{\phantom{0}63}$Eu $\rightarrow$ $^{152}_{\phantom{0}64}$Gd$^*$ + $e^{-}$ + $\overline{\nu}_e$ (27$\%$)\\
\smallskip
\hspace{2cm}$\hookrightarrow$ $^{152}_{\phantom{0}64}$Gd(g.s.) + $\gamma$(779 keV)\\

\end{flushleft}
\end{table*}


\section*{Acknowledgments}
We wish to thank J. J. Coy, R. de Meijer, S. Fancher, J. Herczeg, G. W. Hitt, J. H. Jenkins, D. Koltick, P.A. Sturrock, and T. Ward for helpful communications. We are deeply indebted to the staff of Oak Ridge National Laboratory for their assistance in carrying out this experiment. This study was funded by the Office of Nuclear Energy, U.S. Department of Energy under contract number DE-DT0004091.001b.

\clearpage
\section*{References}

\bibliographystyle{apsrev4-1}
\bibliography{newbiboldlabels.bib}


\end{document}